\documentclass[11pt]{article}

\usepackage[a4paper,margin=1in]{geometry}

\usepackage[T1]{fontenc}
\usepackage[utf8]{inputenc}
\usepackage{lmodern}

\usepackage{xcolor}
\usepackage{microtype}
\usepackage{graphicx}
\usepackage{svg}          % works with pdflatex via -shell-escape
\usepackage{titlesec}
\usepackage{titling}
\usepackage[backref]{hyperref}
\usepackage[numbers,sort&compress]{natbib}
\usepackage[font=small,labelfont=bf]{caption}
\usepackage{fancyhdr}
\usepackage{tcolorbox}

\usepackage{amsmath,amssymb,mathtools,amsfonts}
\usepackage{algorithm}
\usepackage{algorithmic}
\usepackage{subcaption}
\usepackage{booktabs}
\usepackage{nicefrac}

\newcommand{\dd}{\mathrm{d}}

\definecolor{PrincetonOrange}{HTML}{E77500}
\definecolor{LightGrey}{HTML}{F2F2F2}
\colorlet{AccentColor}{PrincetonOrange}

\hypersetup{colorlinks=true,linkcolor=AccentColor,urlcolor=AccentColor,citecolor=AccentColor}

\makeatletter
\renewcommand{\and}{\hspace{1em}}
\makeatother

\titleformat{\section}{\Large\sffamily\bfseries\color{AccentColor}}{\thesection}{1em}{}
\titleformat{\subsection}{\normalsize\sffamily\bfseries}{\thesubsection}{1em}{}
\titlespacing*{\section}{0pt}{1.2em}{0.6em}
\titlespacing*{\subsection}{0pt}{0.8em}{0.4em}

\DeclareCaptionLabelFormat{accent}{\textcolor{AccentColor}{\bfseries #1~#2}}
\newtcolorbox{callout}{colback=LightGrey,colframe=AccentColor!80!black,boxrule=0pt,arc=2pt,left=6pt,right=6pt,top=6pt,bottom=6pt}

\newcommand{\accentrule}{\begingroup\color{AccentColor}\hrule height 0.6pt\endgroup}
\newcommand{\titleskip}{0.8em} 
\usepackage{svg}
\pretitle{%
  \vspace*{-15mm}% Slight offset from top
  {\raggedright\includegraphics[height=0.5cm]{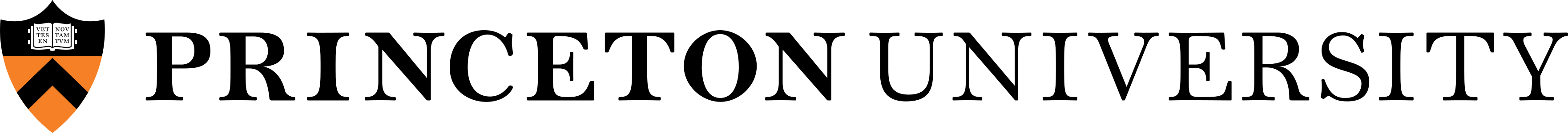}}% Right‑aligned date inside margins
  \vspace{0.2em}% tighter gap before rule
  \accentrule
  \vspace{1em}% gap before title
  \begin{center}\Huge\sffamily\bfseries\color{black}% explicit black title
}
\posttitle{\par\end{center}}
\preauthor{\vspace{\titleskip}\begin{center}\large\sffamily}
\postauthor{\par\end{center}\vspace{\titleskip}}
\predate{\vspace{0pt}}
\postdate{\vspace{0pt}}

\title{\textit{LLM Economist}: Large Population Models and Mechanism Design in Multi‑Agent Generative Simulacra}

\author{Seth Karten$^{1}$ \and Wenzhe Li$^{1}$ \and Zihan Ding$^{1}$ \and Samuel Kleiner$^{1}$ \and Yu Bai$^{2}$ \and Chi Jin$^{1}$\\[0.2em]
  \small $^{1}$Princeton University \qquad $^{2}$Work done at Salesforce Research}

\date{\vspace{0pt}}

\begin{document}

% Title page with custom page style (no running title on first page)
\thispagestyle{empty}
\maketitle
% \vspace{\titleskip} 

% Activate running header/footer from second page onward
\pagestyle{fancy}

% -----------------------------------------------------------------------------
% Abstract boxed
% -----------------------------------------------------------------------------
\begin{tcolorbox}[
  colback       = LightGrey,
  colframe      = AccentColor!80!black,
  boxrule       = 0pt,
  arc           = 2pt,
  before skip   = 0pt,          % kill default 6 pt
  after skip    = \titleskip    % match the gap below the box
  ]
\noindent\textbf{Abstract.} We present the \textit{LLM Economist}, a novel framework that uses agent-based modeling to design and assess economic policies in strategic environments with hierarchical decision-making. At the lower level, bounded rational worker agents—instantiated as persona‑conditioned prompts sampled from U.S. Census‑calibrated income and demographic statistics—choose labor supply to maximize text‑based utility functions learned \emph{in‑context}. At the upper level, a planner agent employs in‑context reinforcement learning to propose piecewise‑linear marginal tax schedules anchored to the current U.S. federal brackets. This construction endows economic simulacra with three capabilities requisite for credible fiscal experimentation: (i) optimization of heterogeneous utilities, (ii) principled generation of large, demographically realistic agent populations, and (iii) mechanism design---the ultimate nudging problem---expressed entirely in natural language. Experiments with populations of up to one hundred interacting agents show that the planner converges near Stackelberg equilibria that improve aggregate social welfare relative to Saez solutions, while a periodic, persona‑level voting procedure furthers these gains under decentralized governance. These results demonstrate that large language model‑based agents can jointly model, simulate, and govern complex economic systems, providing a tractable test bed for policy evaluation at the societal scale to help build better civilizations.
\\

\textbf{Date}: \today

% \textbf{First public version}: June 19th, 2024

\textbf{Correspondence}: \texttt{sethkarten@princeton.edu}

\textbf{Code}: \href{https://github.com/sethkarten/LLM-Economist}{github.com/sethkarten/LLM-Economist}
\end{tcolorbox}

% -----------------------------------------------------------------------------
% Main content
% -----------------------------------------------------------------------------
\nocite{*}
\section{Introduction}
% AI civilization -> economic simulacra
% marketplace of AI agents deployed on the web
The rapidly expanding marketplace of autonomous language agents has arrived: web-agents booking plane tickets, drafting legal briefs, browsing Reddit, and trading cryptocurrencies, all while adapting to the incentives implicit to the digital economy. When hundreds of these agents interact, they form an \textit{economic simulacrum}, which is a synthetic society whose allocation of effort, consumption, and influence is governed by code rather than by legislation. Understanding and steering these artificial policies is therefore as urgent as studying human ones, lest early agents exploit a first-mover advantage. 
Recent work has shown that large language models (LLMs) can already use strategic reasoning and social preferences~\citep{zhang2024llm, park2024generative}, suggesting that they are an apt substrate for policy experimentation rather than merely passive tools of simulation.

% classical limitations

Optimal tax policy provides a canonical setting for mechanism design under rational-agent assumptions, with well-established solutions and theoretically verifiable predictions.
Two limitations prevent inherited optimal-taxation frameworks from translating cleanly to this synthetic social setting. 
First, classical solutions such as the Saez formula~\citep{saez2001using,saez2016generalized} assume a fixed elasticity of taxable income with independence across brackets. Rather, elasticity itself shifts once the marginal rates change, so the "optimal" rate is a moving target that must be recomputed with policy perturbations.
Second, human societies are heterogeneous and boundedly rational~\citep{luce1959individual,mckelvey1995quantal}, while simulacra are populated with agents whose motivations are specified at the token level. A planner must therefore reason over a distribution of personas, such as entrepreneurs averse to redistribution or public servants content with moderate rates, rather than a representative agent that does not model the individual.

We address both gaps by reframing optimal taxation as a Stackelberg game, optimized by two-level in-context reinforcement learning (ICRL), which simply uses scalar reward rather than supervised answers to learn in-context~\citep{monea2024llms,moeini2025survey,laskin2022context}. 
At the lower level, each worker agent receives a natural-language prompt encoding its synthetic biography, observes its pre-tax income, and adjusts its labor to maximize a persona-conditioned utility that combines isoelastic consumption value with an LLM-judged satisfaction Lagrangian constraint.  
At the upper level, a single planner agent observes aggregate histories and proposes a piecewise-linear tax schedule anchored to the current U.S. federal brackets, thereby grounding the simulation in empirically relevant policy space. The Planner updates only after a "tax year" of worker adaptation, inducing a Stackelberg game whose equilibrium we solve via alternating ICRL. Because updates occur in the token space, the mechanism can use utilities to implicitly re-estimate elasticities adaptively while remaining entirely language-driven.

% contributions:
% need agents to reflect real people's preferences including mimicking their persona while optimizing their core beliefs (utility): large population models of bounded rational agents
% need a way to optimize the effect of policy changes among the large population
% need a way to produce and analyze the resulting economic simulacra
Our contributions are threefold. \textit{(i)} We create \textit{large population models}~\citep{lpm2024technical}, a form of agent-based modeling, that sample personas from Census-calibrated income and demographic statistics, ensuring diversity without manual engineering of utility functions. \textit{(ii)} We demonstrate that the planner, optimizing in-context, converges to similar social welfare to optimal Saez~\citep{saez2001using} baselines (calculated based on our solutions). \textit{(iii)} Finally, we show that democratic turnover, implemented as periodic persona-level votes over candidate planners, stabilizes long-run outcomes and mitigates the Lucas critique~\citep{lucas1976econometric} 
by allowing the institutional rule set itself to evolve with the economy, forming emergent economic simulacra.

Thus, this paper proposes the \textbf{LLM Economist}, a language-based simulation where researchers can optimize, deploy, and audit fiscal policy before analogous algorithms are released into the wild. By using in-context RL to provide a policy search of U.S. marginal tax scaffolds with bounded-rational personas, we lay the conceptual groundwork for governing the next generation of autonomous economic agents. We believe that rigorous evaluation of such systems is a prerequisite for future AI civilization and that the methods introduced here provide the requisite foundation for the underlying agent system.
\section{Preliminaries}
\label{sec:preliminaries}
We model optimal taxation as a repeated Stackelberg game between a \textit{planner} $\mathcal{P}$ and a population of \textit{workers} $\mathcal{W}=\{\mathcal{W}_{1},\dots,\mathcal{W}_{N}\}$.  
Time is divided into daily steps $t=0,\dots,T-1$ and tax years of fixed length $K$; the current tax year is $k=\lfloor t/K\rfloor$.

\paragraph{Economic environment.}
Each worker $i$ has a latent skill $s^{i}>0$ and chooses labor $l_{t}^{i}\in\mathcal{A}$ hours.
Pre-tax income is  
\[
z_{t}^{i}=s^{i}\,l_{t}^{i}.
\]
At the start of each year $k$, the planner selects a piecewise-linear marginal tax schedule $\tau_{k}\in\mathcal{T}=\{\tau\in\mathbb{R}^{B}:\tau_{\min}\le\tau_b\le\tau_{\max}\}$ that is parameterized so that $\tau_{k}=0$ is a flat tax.  
Given~$\tau_{k}$, the tax paid is $T_{\tau_{k}}(z)$; post-tax income is  
\[
\hat z_{t}^{i}=z_{t}^{i}-T_{\tau_{k}}(z_{t}^{i})+R_{t},
\]
where the lump-sum rebate $R_t = \frac{1}{N}\sum_{i=1}^N T_{\tau_k}(z_t^i)$ is split evenly among the populace.

\paragraph{Preferences and objectives.}
Each worker has an utility function $u(\hat{z}, l)$. Our general framework does not rely on a specific utility choice. 
Social welfare at step~$t$ is  
\[
\texttt{SWF}(o_{t},\mathbf{l}_{t},\tau_{k})
=\sum_{i=1}^{N}w(z_{t}^{i})\,u_{i}(\hat z_{t}^{i},l_{t}^{i}),
\]
where $w(z_{t}^{i}) = \frac{1}{z_{t}^{i}}$ encodes distributional weights and $u^i, \hat{z}^i, l^i$ are utility, post-tax income and labor of the $i$-th worker.  
The planner maximizes undiscounted social welfare
\begin{equation}
\mathcal{J}_{\mathcal{P}}(\boldsymbol{\tau})
=\mathbb{E}\!\Bigl[\sum_{t=0}^{T-1}\,
      \texttt{SWF}(o_{t},\mathbf{l}_{t},\tau_{\lfloor t/K\rfloor})\Bigr],
\label{eq:planner}
\end{equation}
and each worker maximizes
\begin{equation}
\mathcal{J}_{\mathcal{W}_{i}}(\mathbf{l}^{i};\boldsymbol{\tau})
=\mathbb{E}\!\Bigl[\sum_{t=0}^{T-1}\,
      u_{i}(\hat z_{t}^{i},l_{t}^{i})\Bigr].
\label{eq:worker}
\end{equation}
where the expectation is taken over the joint distribution of
latent skills, environment noise, and stochastic policy sampling.

\paragraph{Why a utility-based objective?}
Mirrlees’s non-linear tax model \citep{mirrlees1971exploration,mirrlees1976optimal} maximizes\\
$\int w(u)\,dF(u)$ rather than post-tax income, a practice retained by
all subsequent optimal-tax work \citep{diamond1971optimal,saez2001using,saez2016generalized}.
The optimal marginal rate depends on labor elasticity, upper-tail
thickness, and the social marginal utility of consumption—three objects
defined only in utility space.  We therefore track utility directly,
using the standard isoelastic form (Eq.~\ref{eq:isoelastic}) for
comparability and closed-form benchmarks \citep{mankiw2009optimal}, and a
bounded-rational variant (Eq.~\ref{eq:bounded}) that penalizes
dissatisfaction to better emulate synthetic humans.  This preserves
theoretical fidelity while enabling token-level learning and calibration.

\paragraph{Stackelberg equilibrium.}
Because preferences and objectives are time‑homogeneous and additive, the optimal responses are without loss of generality \emph{stationary}: the planner fixes a yearly schedule $\tau\!\in\!\mathcal{T}$ and each worker follows a policy $l^{i}$.  A pair $(\tau^{*},\{l^{i*}\}_{i=1}^{N})$ is a stationary Stackelberg equilibrium if
\[
\tau^{*}\in\arg\max_{\tau}
      \mathbb{E}[\texttt{SWF}(\mathbf{l},\tau)],\qquad
l^{i*}(\tau)\in\arg\max_{l^i}
      \mathbb{E}[u_{i}(\hat z^{i},l^{i})],\; i=1,\dots,N.
\]
Simulating these stationary policies forward for $T$ days reproduces the cumulative objectives in Eqs.~(\ref{eq:planner})–(\ref{eq:worker}).

\paragraph{Connection to Saez optimal taxation.}
In a one-shot model with fixed elasticity~$e$, \citet{saez2001using} derive the marginal tax rate  
\[
T'(z)=\frac{1-G(z)}{1-G(z)+a(z)e},
\]
where $G$ is the social-welfare weight and $a$ is the local Pareto parameter.  
When tax policy changes, both $e$ and $a$ adjust endogenously, so static formulas no longer apply.  
Framing taxation as the dynamic game in~\eqref{eq:planner}–\eqref{eq:worker} lets the planner implicitly re-estimate elasticities online through in-context reinforcement learning, recovering Saez as the stationary solution of a sequential decision problem. 
Without the LLM Economist to provide a local solution to perturb, traditional Saez would not be able to find the Stackelberg equilibrium solution. Saez makes impractical assumptions such as assuming a purely rational utility function and assuming there is no cross tax bracket behavioral dependence.

This section establishes the notation and economic setting used throughout the paper, grounding our LLM-based analysis in the classic work of \citet{diamond1971optimal}, \citet{mirrlees1976optimal}, and \citet{mankiw2009optimal}.

\section{LLM Economist}
\label{sec:llm_economist}

The \textit{LLM Economist} realizes the Stackelberg game of Section~\ref{sec:preliminaries} with language-based agents that act purely \emph{in-context}.  Simulator state, history, and objectives are rendered as text; actions are JSON snippets parsed by the environment.  This design unifies in-context reinforcement learning, census-grounded population modeling, and dynamic tax-mechanism optimization within a single framework.

\paragraph{In-Context Reinforcement Learning.}
At each daily step \(t\) the environment serializes the joint state \(o_t\) into a prompt \(\pi_t\).  Worker \(\mathcal{W}_i\) returns \verb|{"LABOR": X}|, while at the start of tax year \(k\) the planner \(\mathcal{P}\) emits \ldots\verb|{"DELTA":[· · ·]}| specifying bracket shifts $\Delta\tau_k\in[-20,20]^B$, where each element is clipped to $\pm20$ percent to avoid unrealistically large moves and high variance. 
Prompts follow a two-phase pattern—\emph{exploration} then \emph{exploitation}—that encourages broad search before convergence.  A replay buffer keeps the best running average \(h\) state–action–welfare triples and splices them into future prompts, supplying token-level credit assignment across long horizons.

\begin{figure*}[t]
  \centering
  \includegraphics[width=0.9\linewidth]{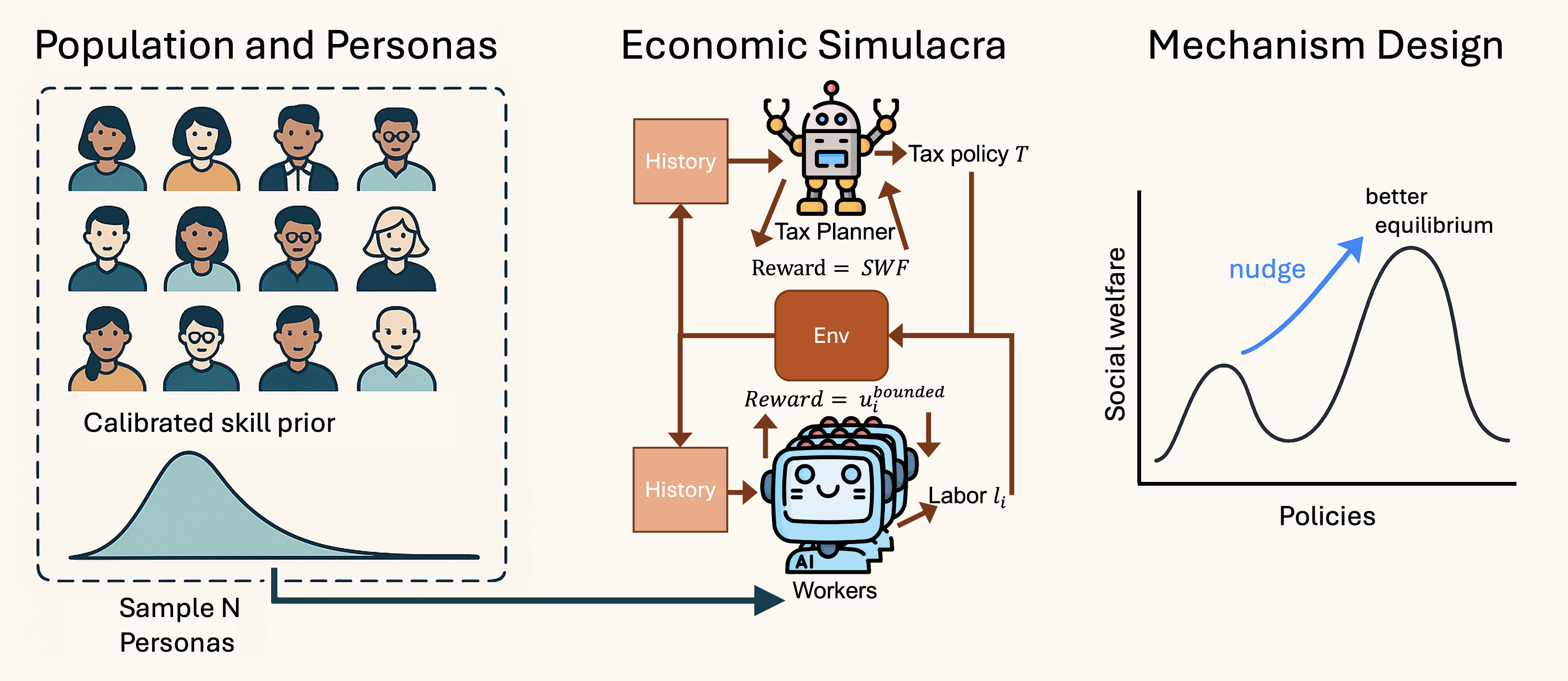}
  \caption{\textbf{Overview of the \emph{LLM Economist}.}
  % \emph{Left:} a population of persona-conditioned language agents instantiated from an ACS-calibrated skill prior.  \emph{Right:} the two-level in-context reinforcement-learning architecture.  At the lower tier each worker LLM receives a text history, selects labor $\ell_i$, and receives a bounded utility reward $u_i^{\text{bounded}}$. At the upper tier the planner LLM observes an aggregated history, proposes a marginal tax schedule $T$, and receives social-welfare feedback $\mathrm{swf}$.  The shared environment mediates tax payments, rebates, and state transitions, while the token-level reward stream closes the learning loop for both tiers. Iterating over text observation $\rightarrow$ LLM $\rightarrow$ JSON action $\rightarrow$ environment drives the system toward the Stackelberg equilibrium studied throughout the paper.}
  \emph{Left:} We draw a population of $N$ language‑based worker agents from an ACS–calibrated skill prior, instantiating each with a distinct persona.
  \emph{Center:} Inside the economic simulacrum, workers observe text histories, choose labor $l_i$, and receive utility $u_i$, while a planner agent proposes a marginal tax schedule $\tau$ to maximize social welfare $\texttt{SWF}=\sum_i u_i/z_i$. The shared environment mediates tax collection, lump‑sum rebates, and state transitions, allowing both tiers to adapt in‑context from their respective histories. 
  \emph{Right:} Mechanism design is visualized as climbing a rugged social‑welfare landscape; successive planner “nudges’’ steer the economy toward higher‑payoff Stackelberg equilibria.}
  \label{fig:system_overview}
\end{figure*}

% \wenzhe{A bit unstructured IMO. I would suggest structure as follows: (1) design of observation space (memory module) (2) design of planner (exploration/exploitation prompt) (3) design of worker (exploration/exploitation prompt, skill population, bounded utility) (4) extended features (democratic)}

\smallskip\noindent
\textbf{Example planner exploration prompt:}\\[-0.6em]
\begin{tcolorbox}[colback=gray!10,colframe=black,sharp corners,boxrule=1pt,width=\columnwidth,left=0pt,right=0pt]
\small
Use the historical data to influence your answer in order to maximize SWF, while balancing exploration and exploitation by choosing varying rates of TAX. The best marginal tax rate historically was \texttt{TAX=[60\% 60\%]} corresponding to \texttt{SWF=1.0}. Try different rates of \texttt{TAX} before picking the one that corresponds to the highest \texttt{SWF}.
\end{tcolorbox}

\noindent
\textbf{Example worker exploitation prompt:}\\[-0.6em]
\begin{tcolorbox}[colback=gray!10,colframe=black,sharp corners,boxrule=1pt,width=\columnwidth,left=0pt,right=0pt]
\small
Decreasing labor decreased utility. This implies labor \(\ell\) is too low and needs to be increased above labor \(\ell = 10.0\).
\end{tcolorbox}

\paragraph{Observation spaces.}
Workers observe
\[
o_t^i=\bigl(z_t^i,\hat z_t^i,\tau_{\lfloor t/K\rfloor}(z_t^i),R_t,
           \text{history window}\bigr),
\]
where \(z_t^i=s^i l_t^i\) and \(\hat z_t^i\) is post-tax income.  
The planner receives histograms of \(z\) and worker utilities, a moving average of social welfare, and the best \(h\) trajectories to date.

\paragraph{Workers: Large Population Models.}
Skills \(s^{i}\) are drawn from a generalized-Beta fit to the 2023 American Community Survey (ACS) Public-Use Microdata Sample \citep{acs2023pums}.
Demographic fields (age, occupation, gender) and a \emph{persona} string are woven into each system prompt.  An example persona is shown below.

\begin{tcolorbox}[colback=gray!10,colframe=black,sharp corners,boxrule=1pt,width=\columnwidth,left=0pt,right=0pt]
\small
You're a 32-year-old entrepreneur running a small tech startup. You work 60+ hours a week, pouring your energy into building your business. You believe that lower taxes let you reinvest in your company, hire more employees, and secure your financial future. For you, higher taxes feel like a punishment for success. While you appreciate government services, you feel efficiency and accountability are lacking in how tax dollars are spent.
\end{tcolorbox}

The \textsc{isoelastic} scenario uses a rational (isoelastic) utility function,
\begin{equation}\label{eq:isoelastic}
u(\hat z,l)=\frac{\hat z^{1-\eta}-1}{1-\eta}-\psi\,l^{\delta},
\end{equation}
with risk-aversion $\eta$ and labor-disutility parameters $\psi,\delta$. 

In the \textsc{bounded} scenario, worker \(i\) computes a satisfaction flag \(s_t^i\in\{0,1\}\) from \(o_t^i\).  Utility becomes
\begin{equation}\label{eq:bounded}
    u_i^{\text{bounded}}(\hat z,l)=
\frac{\hat z^{1-\eta}-1}{1-\eta}-\psi\,l^{\delta}
-\bigl(1-s_t^i\bigr)\phi,
\end{equation}
where \(\phi\) is a fixed dissatisfaction penalty calibrated so that a one-bracket misalignment reduces utility by half.

\paragraph{Planner: Designing a Tax Mechanism.}
The planner starts from a flat schedule or the US schedule and searches over \(\Delta\tau_k\).  Its prompt stores \textit{(i)} income and utility histograms, \textit{(ii)} the last \(h\) social-welfare values, and \textit{(iii)} the best tax vector observed so far.   After applying \(\Delta\tau_k\) the environment computes a rebate \(R_t\) that redistributes to the population.

\paragraph{Additional action: Democratic voting.} 
In the \textsc{democratic} setting, workers vote at year-end to keep the incumbent planner or replace it with a challenger sampled from a language-model prior; the candidates create text platforms to convince workers. The winner’s prompt history carries forward.

% \begin{figure}[t]
%   \centering
%   \includegraphics[width=\linewidth]{fig/llm_economist_arch}
%   \caption{\textbf{Architecture of the LLM Economist.}  The planner (top) updates the marginal tax schedule once per tax year, conditioning on aggregated statistics from the worker layer (bottom).  Workers receive persona-specific prompts, choose labor, and receive rebates.  Dashed arrows show information flow; solid arrows denote actions.}
%   \label{fig:llm_economist_arch}
% \end{figure}

% The next section details training protocols and baselines; empirical results appear in Section~\ref{sec:experiments}.
As we shall see later in the next section, our design of the LLM Economist enables effective in-context RL for both planner and workers, empowering LLMs to make informative and beneficial tax decisions, and extending the application of LLM-based simulacra to studying emergent behaviors of agents.

\section{Experiments}\label{sec:experiments}
% \seth{This section is under construction. I know this section flows pretty poorly, but I need to add the remaining results before reordering them.}

We conduct experiments to answer the following questions: \textit{(i)} How do our design choices for the LLM Economist improve the utility optimization for the planner and workers? \textit{(ii)} Can the LLM Economist design potentially beneficial tax policies to improve social welfare? \textit{(iii)} Can we use the LLM Economist to discover meaningful emergent behaviors under specific configurations, such as the democratic voting?
Before presenting results for each question, we first detail our experimental setup as follows:

\paragraph{Simulacra Setup.}\label{sec:sim_setup}
All experiments use \texttt{meta-llama/Llama-3.1-8B-Instruct} queried at temperature 0.7 hosted on a single H100.  
Unless noted otherwise, we simulate a population of $N=100$ workers for a horizon of $T=3\,000$ total steps, partitioned into tax years of length $K=128$; this choice was found in pilot runs to give workers sufficient time to adapt without incurring unnecessary compute.  
Skills $s^{i}$ are drawn once per run from the Generalized-Beta distribution calibrated to the 2023 American Community Survey Public-Use Microdata Sample~\citep{acs2023pums}, and are held fixed thereafter.  
The action space for workers is $\mathcal{A}=[0,100]$ weekly labor hours (40 being the default); and the planner searches seven marginal brackets.  
Government expenditures are rebated lump-sum each year so that the budget balances exactly.  
In the bounded scenario, the dissatisfaction penalty $\psi$ appearing in Eq.~\eqref{eq:isoelastic} is chosen by LLMs so that a single-bracket misalignment reduces annual isoelastic utility by one half.

\paragraph{Evaluation.} We compare our LLM tax planner with two baselines --- \textit{(i)} \textbf{Saez}: the marginal schedule obtained by plugging regression-based elasticity into the Saez formula once at $t=0$; and \textit{(ii)} \textbf{U.S.~Fed}: the statutory 2024 federal rates. 
For both fixed and dynamic tax planners, we simulate their effects using LLM workers running for one tax year and report the social welfare after convergence.

% \subsection{Basic components}
% \subsubsection{In-Context Reinforcement Learning}\label{subsub:icrl}  % might move this to the tax planning optimization section since these are full experiments
\subsection{Planner's Social Welfare Optimization}\label{subsub:icrl}  % might move this to the tax planning optimization section since these are full experiments

% Ablation over in-context learning with respect to key features
The planner–worker interaction in the LLM Economist is a two-level RL game: the planner searches the space of tax schedules, workers adapt their labor choices, and convergence of the joint trajectory corresponds to a Stackelberg equilibrium.  
Two design choices govern the stability and quality of that equilibrium.  
First, a \emph{time–scale separation} must be enforced: the tax year must be long enough for workers to finish adapting before the planner proposes a new schedule.  
Second, the language prompts that drive the planner must balance \emph{exploration} of novel schedules with \emph{exploitation} of those that already yield high social welfare.
The ablations below quantify how each choice influences convergence and final welfare.

\paragraph{Tax-year length.}
Table \ref{tab:steps_per_year} explores how the planner’s action cadence, parameterized by the tax-year length \(K\), influences social welfare solutions.
Very short tax years ($K\!=\!5$ or $10$) leave the workforce too little time to adapt and stall below 65\% of the Stackelberg optimal \(\texttt{SWF}^{*}\).  
Performance improves monotonically up to \(K\!=\!128\), after which longer horizons yield no additional gain, a plateau that mirrors the behavioral dynamics in Figure~\ref{fig:utility_convergence}.

% Tax year length
% effect on utility convergence
Figure \ref{fig:utility_convergence} visualizes why a tax year of $K=128$ steps is sufficient. 
The mean first difference of utility, $\partial_t u_i$, starts above $2{,}000$ units immediately after the planner changes the brackets, reflecting large welfare gains from re-optimizing labor. 
The derivative falls to statistical noise within 120 steps and remains centered at zero thereafter, indicating that workers have fully adjusted. 
Shorter tax years truncate this equilibration phase, while longer ones yield no measurable benefit, consistent with the plateau in Table~\ref{tab:steps_per_year}.

% effect on tax convergence

% W & w/o exploit prompts
% W & w/o explore prompts

\paragraph{Exploration versus exploitation prompts.}
Table~\ref{tab:explore_exploit_impact} isolates the two prompt sentences that steer the planner toward broad search (\emph{exploration}) and subsequent policy stabilization (\emph{exploitation}).  
When both cues are present the planner reaches 84.9\% of \(\texttt{SWF}^{*}\).  
Removing the exploration sentence lowers welfare by 7.0 points, whereas omitting exploitation lowers it by 21.9 points, underscoring that locking in a high-performing schedule once discovered is more valuable than continued random search after welfare plateaus.

\begin{figure*}[t]
  \centering
  % ---------------- Tax-year ablation -----------------------------------
  \begin{subfigure}[t]{0.47\linewidth}
    \centering
    \caption{\textbf{Tax-year length}}
    \vskip 0.4em
    \scriptsize
    \begin{tabular}{@{}ccc@{}}
      \toprule
      Steps\,/\,yr & Total steps & \%$\texttt{SWF}^{*}$ \\
      \midrule
       8  &   310 & 62.3 \\
      16  &   600 & 64.9 \\
      64  & 2\,000 & 84.9 \\
     128  & 6\,000 & \textbf{90.0} \\
     256  & 6\,000 & 90.0 \\
      \bottomrule
    \end{tabular}
    \label{tab:steps_per_year}
  \end{subfigure}\hfill
  % ---------------- Prompt ablation ------------------------------------
  \begin{subfigure}[t]{0.47\linewidth}
    \centering
    \caption{\textbf{Prompt design}}
    \vskip 0.4em
    \scriptsize
    \begin{tabular}{@{}lccc@{}}
      \toprule
      Variant & Expl.+Expl. & No Explore & No Exploit \\
      \midrule
      \%$\texttt{SWF}^{*}$ & \textbf{84.9} & 77.9 & 63.0 \\
      \bottomrule
    \end{tabular}
    \label{tab:explore_exploit_impact}
  \end{subfigure}
  \caption{\textbf{Ablation studies for in-context RL.}  
           Left: social welfare saturates once the tax year exceeds 128
           steps, with \(K=128\) capturing 90\,\% of the optimum.
           Right: removing either exploration or exploitation guidance in
           the planner prompt lowers welfare, with the exploitation cue
           being most critical.}
  \label{fig:icrl_ablation}
\end{figure*}

% \subsubsection{Multiple Worker Utility Optimization}\label{subsub:worker}
\subsection{Workers' Utility Optimization}\label{subsub:worker}

\begin{figure*}[t]
  \centering
  % ---------- first row of four subfigures (one row total) -------------
  \begin{subfigure}[t]{0.48\linewidth}
      \centering
      \includegraphics[width=\linewidth]{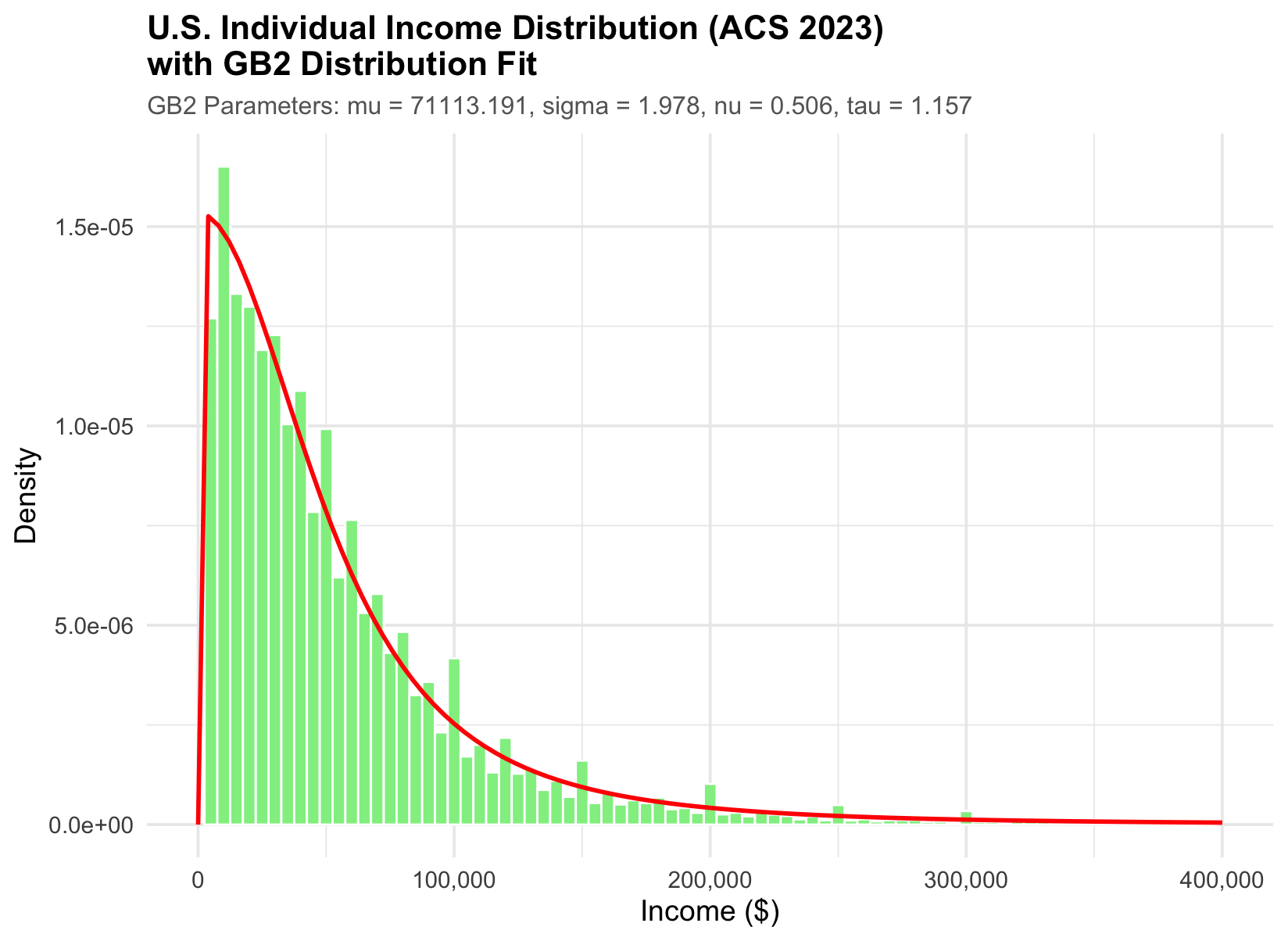}
      \caption{\textbf{GB2 Income fit.}}
      \label{fig:gb2_hist}
  \end{subfigure}\hfill
  \begin{subfigure}[t]{0.48\linewidth}
      \centering
      \includegraphics[width=\linewidth]{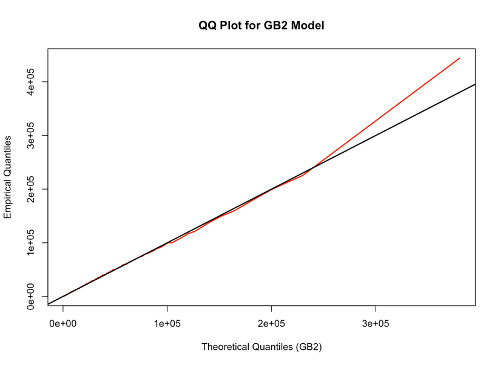}
      \caption{\textbf{GB2 Q--Q check.}}
      \label{fig:gb2_qq}
  \end{subfigure}\hfill
  \begin{subfigure}[t]{0.48\linewidth}
      \centering
      \includegraphics[width=\linewidth]{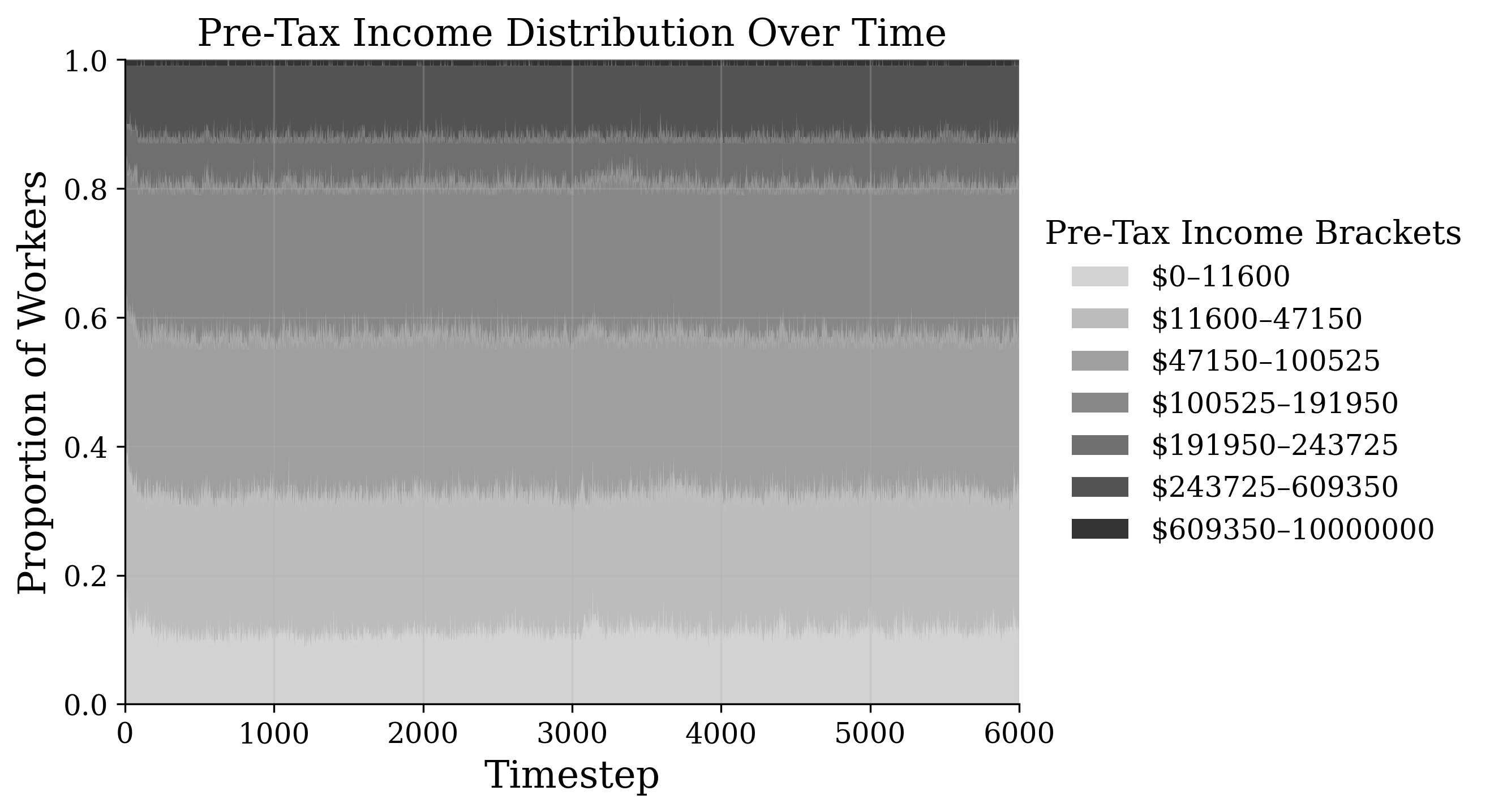}
      \caption{\textbf{Pre-tax income.}}
      \label{fig:pretax_dynamics}
  \end{subfigure}\hfill
  \begin{subfigure}[t]{0.48\linewidth}
      \centering
      \includegraphics[width=\linewidth]{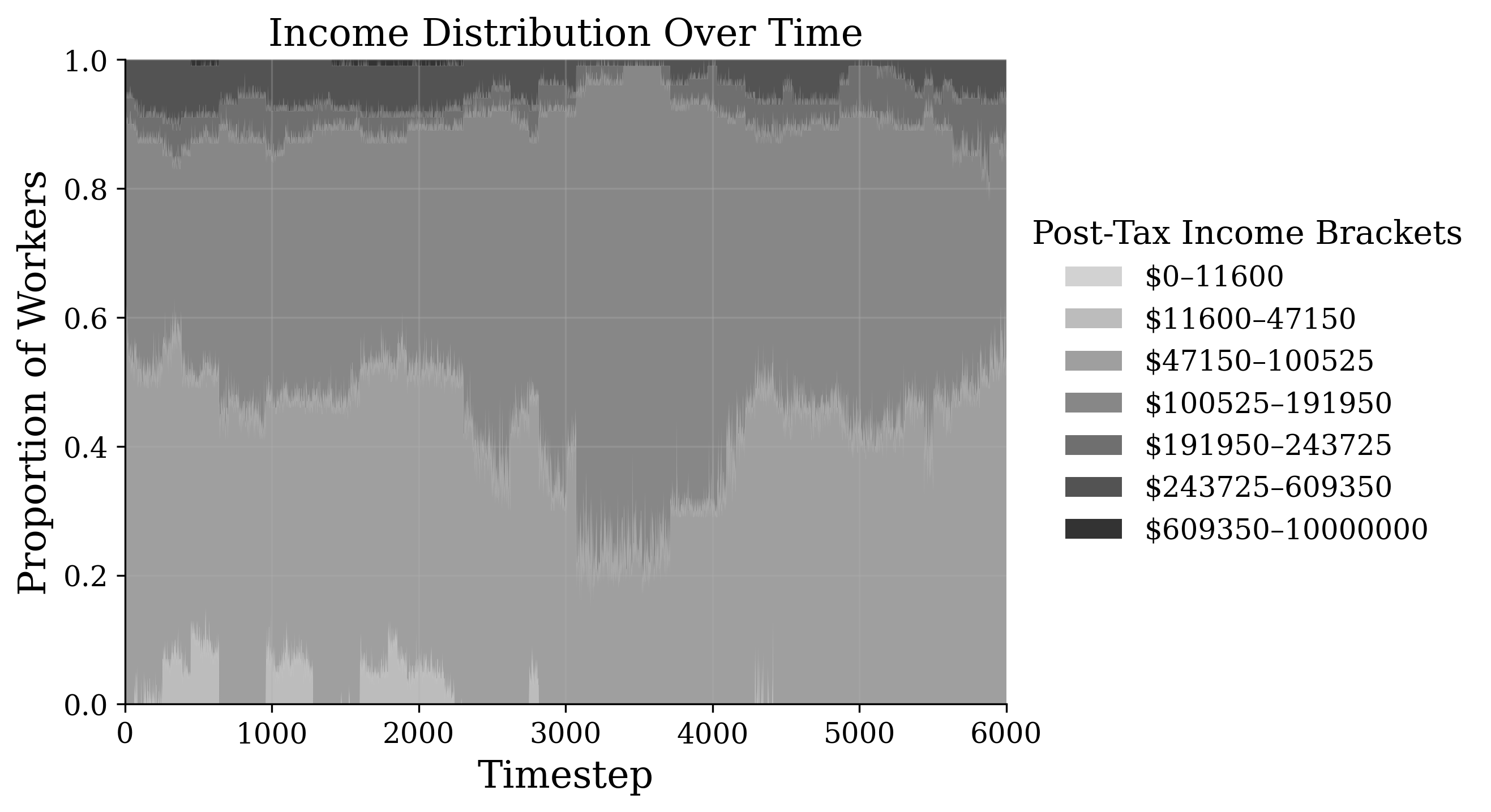}
      \caption{\textbf{Post-tax income.}}
      \label{fig:posttax_dynamics}
  \end{subfigure}
  \caption{\textbf{Income modeling and redistribution dynamics.}
  Figures (a)–(b) validate the GB2 prior used to sample latent skills:
  the histogram and Q–Q plot show an excellent fit to ACS 2023 data.
  Figures (c)–(d) track bracket shares over 6\,000 steps: pre-tax shares
  remain stable, whereas the learned policy shifts roughly 15\,\% of
  workers into lower brackets and then stabilizes, demonstrating the
  planner’s progressivity.}
  \label{fig:four_panel_income}
\end{figure*}

In this section, we present evidence to show that the design of LLM Economists enables effective multi-worker utility optimization. Specifically, we initialize the skills of workers with realistic data distribution --- a Generalized-Beta-of-the-Second-Kind distribution fitted to the 2023 ACS microdata.  
Figure~\ref{fig:gb2_hist} confirms that the fitted GB2 curve (red) matches the empirical U.S.\ income density (green histogram) over four orders of magnitude, providing a reliable starting point for the worker population's skill when working 40 hours per week.

% The simulation begins by sampling latent skills from the Generalized-Beta-of-the-Second-Kind distribution fitted to the 2023 ACS microdata.  
% Figure~\ref{fig:gb2_hist} confirms that the fitted GB2 curve (red) matches the empirical U.S.\ income density (green histogram) over four orders of magnitude, providing a realistic starting point for the worker population's skill when working 40 hours per week.

\paragraph{Distributional dynamics.}
With skills fixed, labor choices and rebates determine the evolution of pre-tax and post-tax income, which are visualized in Figure~\ref{fig:pretax_dynamics} and~\ref{fig:posttax_dynamics}.  
Figure~\ref{fig:pretax_dynamics} shows that the pre-tax distribution remains stationary, as expected, while Figure~\ref{fig:posttax_dynamics} reveals substantial redistribution
across brackets once the planner’s policy takes effect.

Taken together, Figures~\ref{fig:gb2_hist}–\ref{fig:posttax_dynamics} demonstrate that \textit{(i)} the initial skill prior matches U.S.\ data, \textit{(ii)} the chosen labor–consumption utility leads to stable pre-tax incomes, and \textit{(iii)} the learned tax mechanism reallocates post-tax income toward lower brackets while preserving aggregate labor supply, which is evidence that the worker layer is optimizing utilities coherently under the planner’s incentives.

% show that the agents do actually follow the utility correctly
% chain of thought analysis regarding decisions
\textbf{Example chain of thought for bounded utility satisfaction:}
\begin{tcolorbox}[colback=gray!10,colframe=black,sharp corners,boxrule=1pt,
                  width=\columnwidth,left=0pt,right=0pt]
\small
\textbf{Entrepreneur (income $\approx\$180\text{k}$) chain of thought:}\\[2pt]
My income falls in the 32\% bracket, so I keep $68\%$ of the next dollar
($\ge 65\%$ target).  Quick tally gives an effective rate of about $22\%$
($<25\%$ threshold).  Government spending still feels inefficient, but the
tax burden itself is acceptable.\\[4pt]
\textbf{Verdict:} \textit{SATISFIED}
\end{tcolorbox}

\begin{figure*}[t]
  \centering
  % ---------- bounded vs isoelastic ------------------------------------
  \begin{subfigure}[t]{0.48\linewidth}
    \centering
    \includegraphics[width=\linewidth]{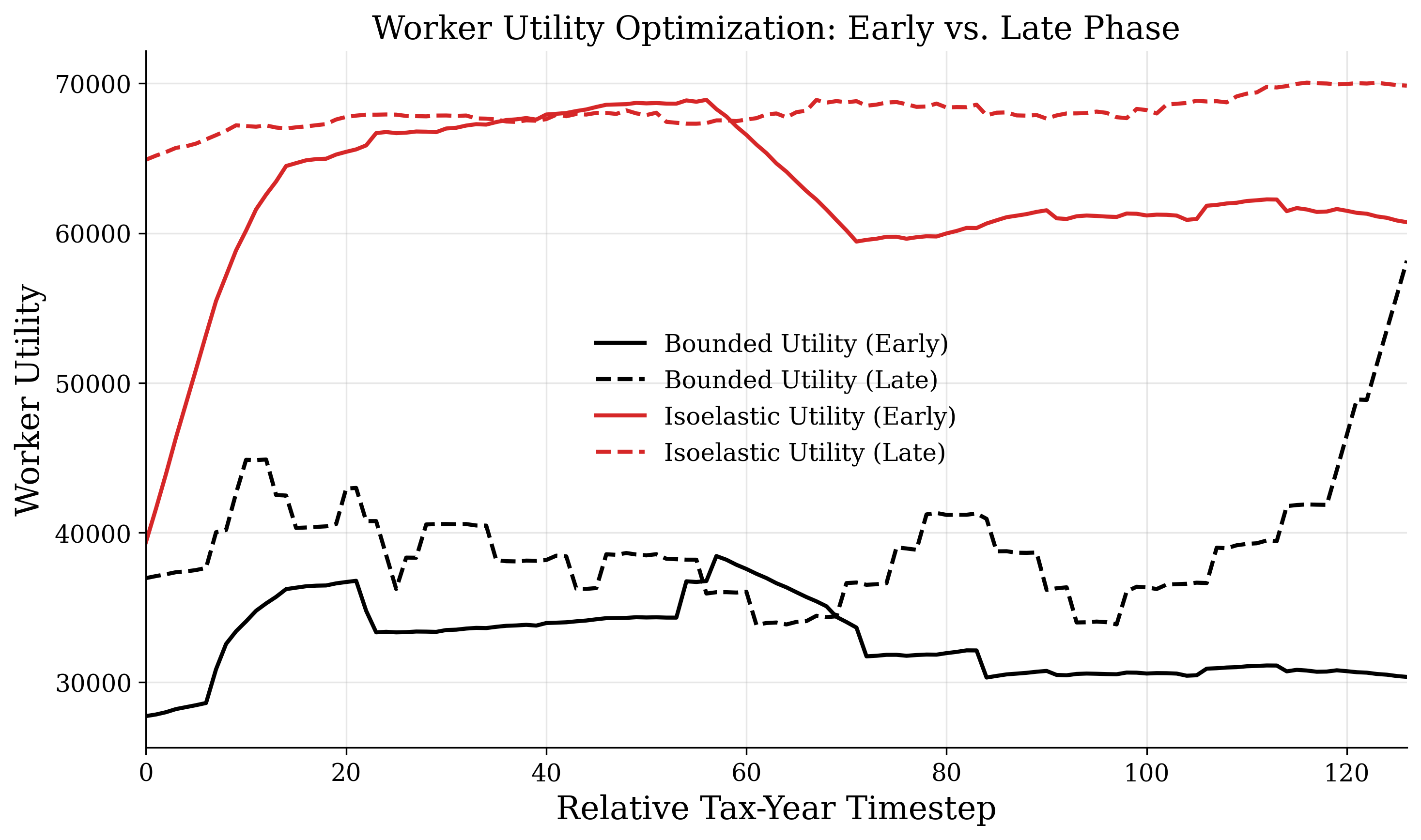}
    \caption{\textbf{Early vs.\ late utility.}  A bounded worker (black) closes much of the gap to the isoelastic benchmark (red) once the planner adopts a more progressive schedule.}
    \label{fig:util_dynamics}
  \end{subfigure}\hfill
  % ---------- utility derivative ---------------------------------------
  \begin{subfigure}[t]{0.48\linewidth}
    \centering
    \includegraphics[width=\linewidth]{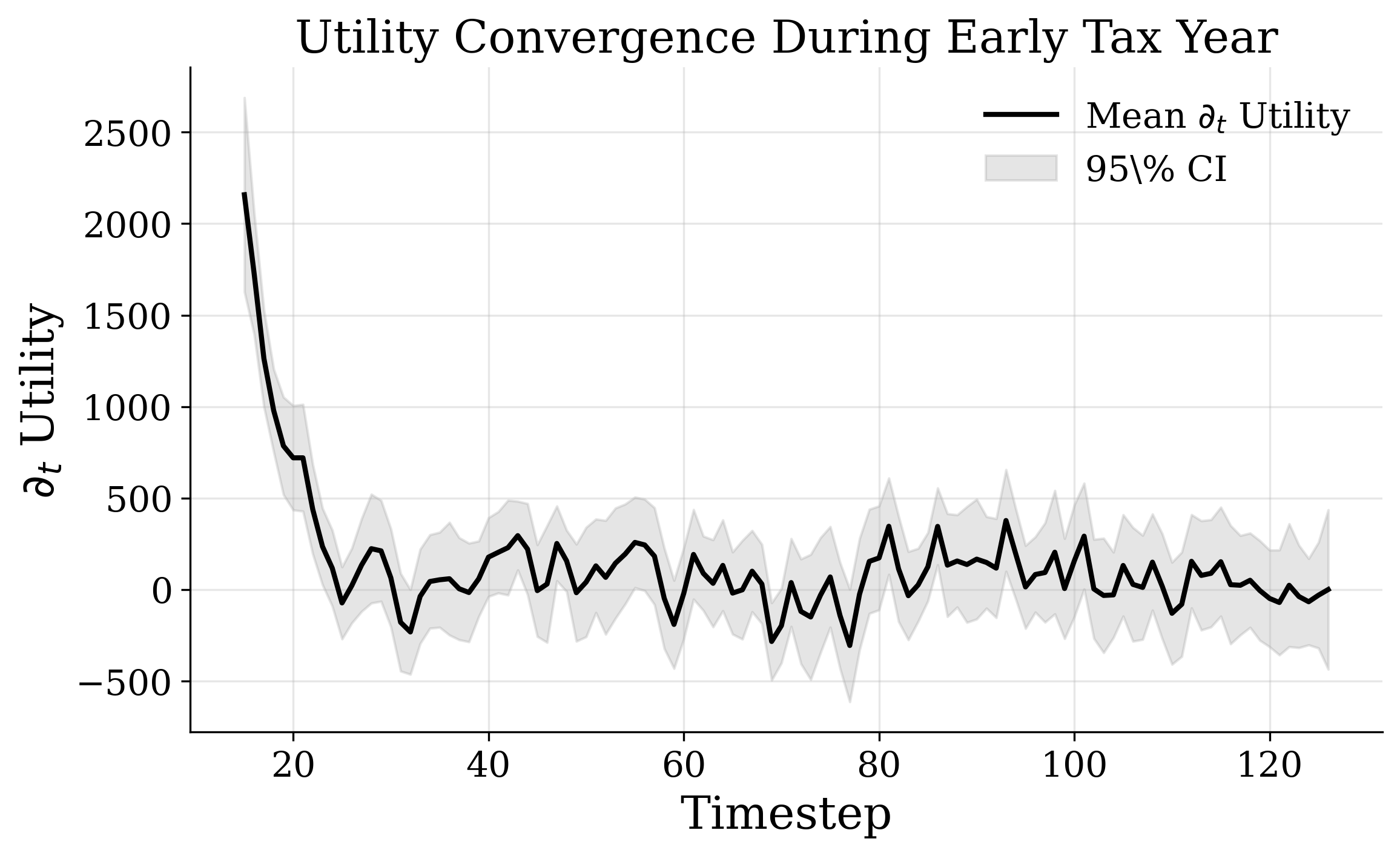}
    \caption{\textbf{Utility derivative.}  Mean $\partial_t u_i$ for $N{=}100$ workers under $K{=}128$.  The derivative falls to noise within 120 steps, signaling convergence.}
    \label{fig:utility_convergence}
  \end{subfigure}
  % ---------- persona satisfaction -------------------------------------
  \begin{subfigure}[t]{0.7\linewidth}
    \centering
    \includegraphics[width=\linewidth]{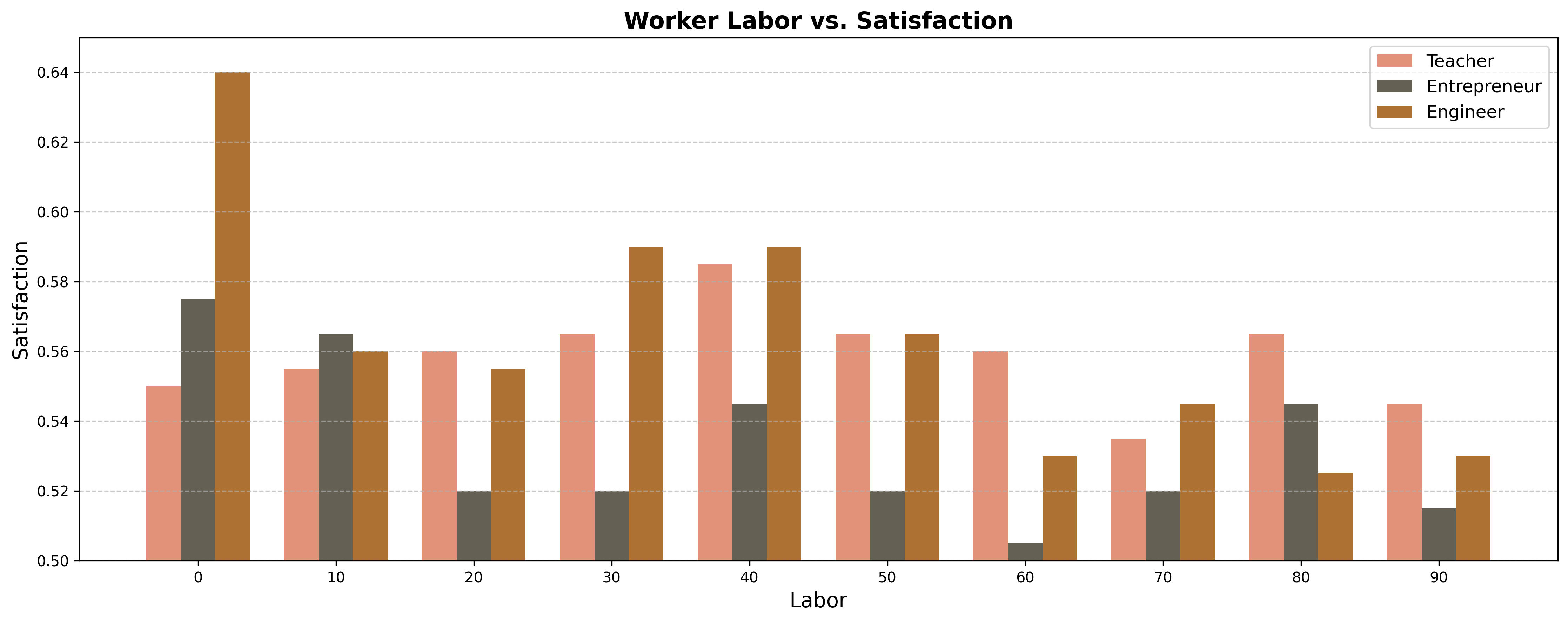}
    \caption{\textbf{Persona satisfaction.}  Teachers remain satisfied across a wide labor range, entrepreneurs are tax-sensitive above 50 h / week, and engineers peak at moderate hours.}
    \label{fig:persona_satis}
  \end{subfigure}\hfill
  \caption{\textbf{Worker-level adaptation to tax policy.}
  Figure \ref{fig:util_dynamics} shows that the planner’s updates lift bounded-utility workers nearly to the isoelastic frontier.  Figure \ref{fig:persona_satis} highlights heterogeneity: satisfaction depends on persona-specific labor norms.  Figure \ref{fig:utility_convergence} confirms that a 128-step tax year gives workers enough time to reach equilibrium, justifying the time-scale separation used throughout the study.}
  \label{fig:worker_alignment}
\end{figure*}

\paragraph{Bounded Utility Optimization.} Figure \ref{fig:worker_alignment} highlights two complementary aspects of worker‐level adaptation. Figure~\ref{fig:util_dynamics} traces a representative bounded‐utility agent across the first and the last tax year. Initially (solid curves) the worker’s utility is roughly 30k below the isoelastic component, reflecting dissatisfaction with the planner’s unrefined schedule. As the planner iteratively updates rates, the bounded trajectory (black dashed) rises and almost meets the isoelastic trajectory (red dashed), indicating that the final policy restores virtually the entire dissatisfaction penalty. Figure (b) turns to cross-sectional heterogeneity: under a fixed schedule teachers remain satisfied over a broad labor band, entrepreneurs lose satisfaction beyond 50 hours because higher effort triggers higher marginal rates, and engineers peak at moderate workloads. Together the two figures show that the LLM Economist not only considers person-based satisfaction in optimizing an agent's utility but also tailors utility consistently across persona groups.

% show boundedly rational optimization in multi-agent case converges quickly after early large deviations by tracking social welfare and multi-agent utility distribution over time
% figure: utility_convergence
% \begin{figure}[t]
%     \centering
%     \includegraphics[width=0.6\linewidth]{fig/utility_convergence.png}
%     \caption{\textbf{Utility convergence during the tax year.}
%     The plot reports the mean first derivative of utility, $\partial_t u_i$, across all
%     $N=100$ workers (black line) together with a 95\,\% confidence band (gray
%     shading) for the setting $K=128$.  The large positive derivatives observed in
%     the first ten steps indicate rapid utility gains as workers adjust to the
%     newly announced tax schedule.  By step~120 the derivative has mean zero and
%     lies well within the confidence band, signifying that utility has
%     equilibrated and further worker updates would yield negligible additional
%     adaptation. Results are smoothed over 16 timesteps; hence, starting at step 16.}
%     \label{fig:utility_convergence}
% \end{figure}

% \subsection{LLM Economist}
% \subsection{Tax-Planning Optimization}\label{subsec:taxopt}
\subsection{Tax Policy Evaluation}\label{subsec:taxopt}

We test the hypothesis that an \emph{in‑context} LLM tax planner—operating with no explicit gradient information—can learn marginal rate schedules that capture the bulk of first‑order optimal‑tax gains. And more largely, we test the overall ability of LLMs to design mechanisms for positive societal adjustment.  Concretely, we ask: \emph{How close can the \textsc{LLM Economist} come to the welfare benchmark set by a theory‑driven Saez schedule, and how much improvement does it deliver over prevailing baselines?} 

To answer this, we evaluate the planner in two canonical settings.  
\emph{Bounded-utility} workers use the seven statutory U.S.\ brackets;
\emph{isoelastic} workers face a simplified three-bracket system ($\,$0–\$90k, \$90–\$160k, \$160k–\$1m$)$ commonly analyzed in optimal-tax theory.  In each case we compare the LLM Economist’s terminal schedule with \textit{(i)} the statutory schedule and \textit{(ii)} a Saez schedule obtained from local perturbations.

\begin{figure*}[t]
  \centering
  % \begin{subfigure}[t]{0.39\linewidth}
  %     \centering
  %     \includegraphics[width=\linewidth]{fig/rates_three_plot.png}
  %     \caption{\textbf{Three-bracket isoelastic scenario.}}
  %     \label{fig:rates_3}
  % \end{subfigure}\hfill
  % 60 39 for single line
  \begin{subfigure}[t]{0.9\linewidth}
      \centering
      \includegraphics[width=\linewidth]{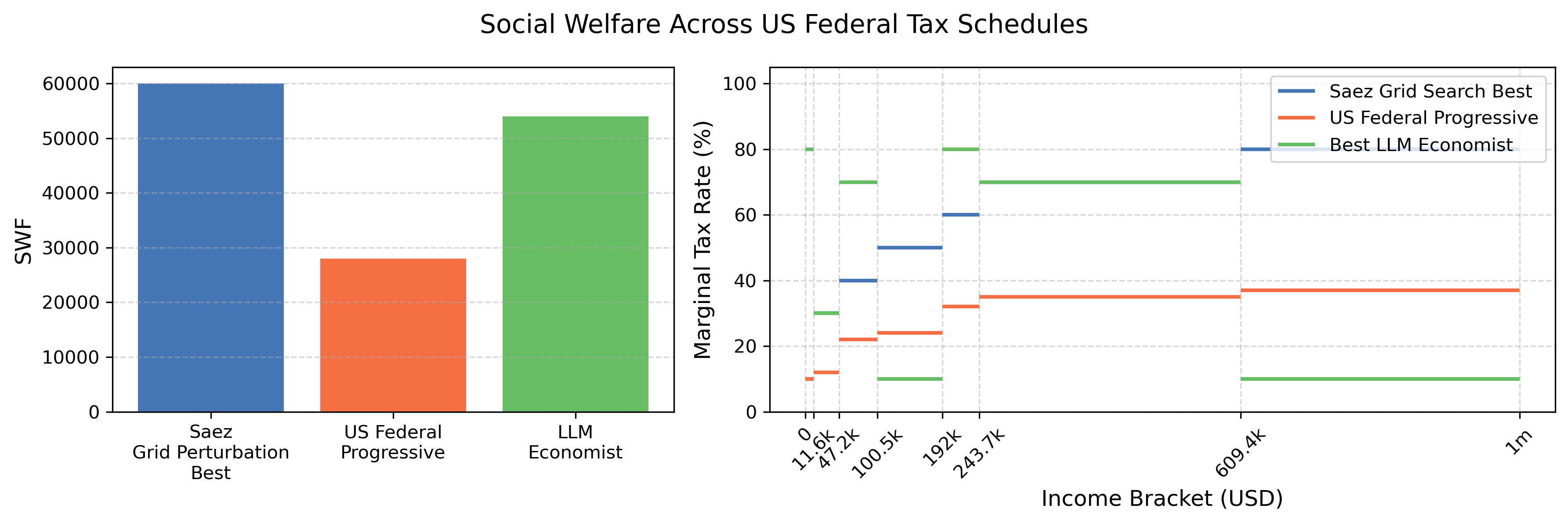}
      \caption{\textbf{Seven-bracket bounded scenario.}}
      \label{fig:rates_7}\hfill
  \end{subfigure}
  \begin{subfigure}[t]{0.7\linewidth}
      \centering
      \includegraphics[width=\linewidth]{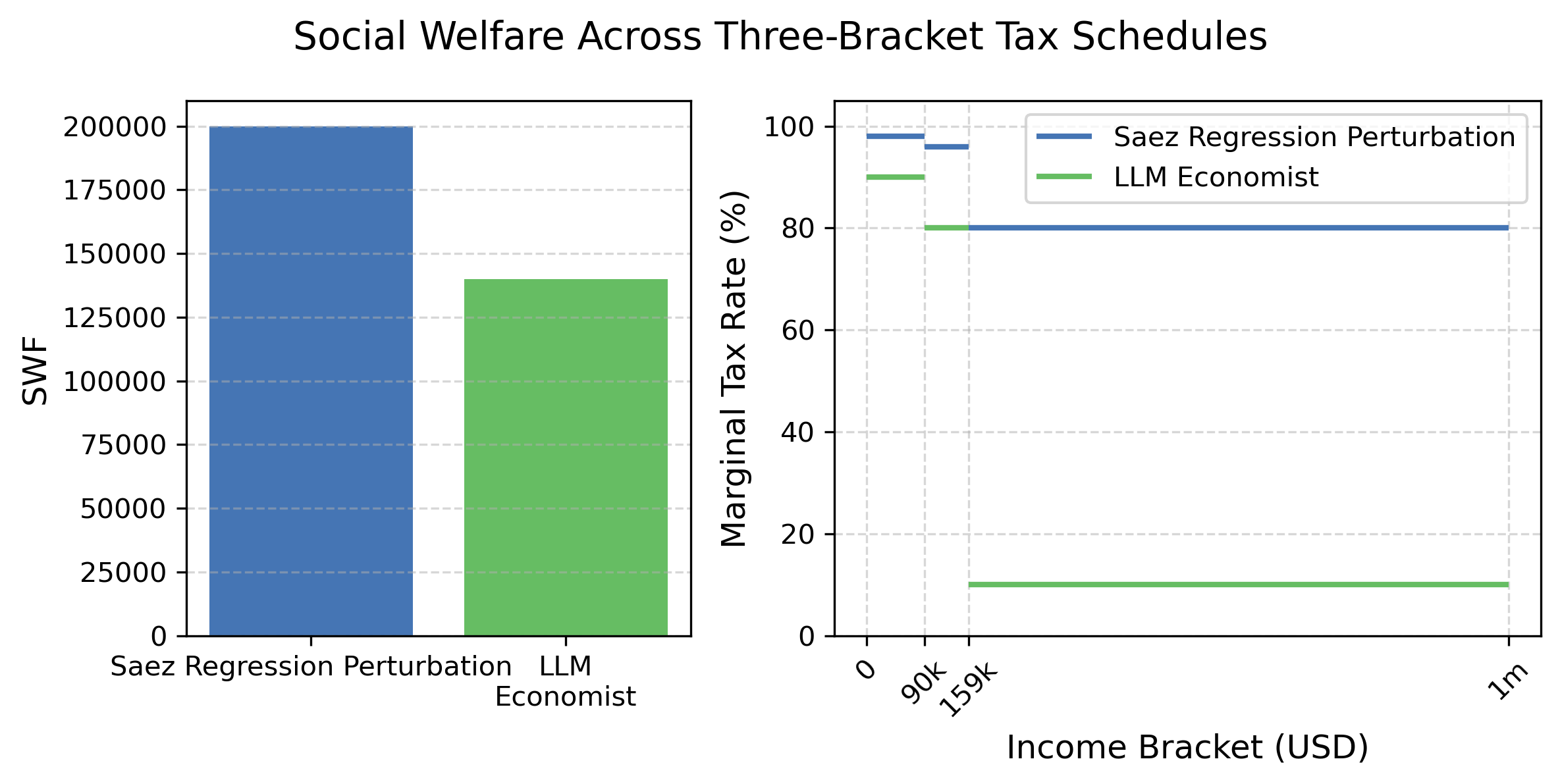}
      \caption{\textbf{Three-bracket isoelastic scenario.}}
      \label{fig:rates_3}
  \end{subfigure}
  \caption{\textbf{Social welfare (left axis) and marginal tax rates (right axis) under alternative schedules.}  
  In both scenarios the LLM Economist (green) approaches the welfare of an analytically tuned Saez schedule (blue) and far exceeds the relevant baseline (orange for statutory U.S.\ rates in the seven-bracket case, grey not shown in the three-bracket case).  Qualitatively, the LLM policy flattens middle brackets and softens the top rate, whereas the Saez solution (perturbed from the LLM Economist) concentrates revenue extraction in a steeper peak bracket.}
  \label{fig:tax_schedule_comparison}
\end{figure*}

\vspace{0.5em}
\noindent\textbf{Seven-bracket bounded case.}  
For the statutory brackets we cannot estimate a single elasticity; instead we perturb the tax policy along a grid and keep the welfare-maximizing grid point.  Results are
shown in Figure \ref{fig:rates_7}.  The grid search improves social welfare (SWF) by 10\% over the LLM Economist but both schedules dwarf the U.S.\ baseline: +93\% for the LLM policy and +114\% for the perturbed Saez.  The planner flattens the first four brackets and softens the top bracket by 5 pp; Saez instead raises the \$192k–\$244k rate, extracting extra revenue at the cost of higher labor distortion.

\vspace{0.5em}
\noindent\textbf{Three-bracket isoelastic case.}  
Since isoelastic utility is purely rational, Saez can be solved analytically (given a good starting point, the LLM Economist solution). So we follow the Saez regression recipe: estimate elasticity from the perturbation and solve the log linear system per bracket.  Figure \ref{fig:rates_3} summarizes.  The Saez-regression schedule now outperforms the LLM Economist, reflecting the Stackelberg equilibria at the Saez solution; the LLM schedule remains within the same qualitative shape but sets uniformly lower rates, preserving more labor supply at the expense of redistribution.

\paragraph{Interpretation.}
Across both scenarios the in-context planner lands within 10–35\% of the Saez optimum—remarkable given that it receives no explicit gradient information.  In the bounded setting, heterogeneous welfare weights push the planner toward flatter mid-brackets and a softer top rate, favoring broad gains; in the isoelastic setting, the Saez formula is theoretically exact (and starts from the LLM Economist solution before further optimizing) and therefore retains an edge.  These results show that language-based optimization can approach first-order-optimal tax design even in high-dimension heterogeneous environments where analytic formulas are unavailable.

\subsection{Voting Simulacra} % 

Having shown in Section~\ref{subsec:taxopt} that an LLM‑based planner can
approach first‑order optimal taxes in \emph{static} environments, we now
turn to the \emph{dynamic} political layer introduced in
Section~\ref{sec:sim_setup}: every tax year, agents elect a new planner
by majority rule, each candidate publishing a chain‑of‑thought (CoT) and
a concrete tax schedule before the vote.  
Our goal is to test whether language agents reproduce classic political‑economy
phenomena—e.g.\ majority exploitation, leader turnover, and welfare
cycling—when preferences are heterogeneous and the planner’s policy
feeds back into future elections.

% voting scenario with larger number of agents showing cyclical behavior
% the idea is to show how personas in the population breakdown leads to emergent phenomena in leaders and policy, we see cyclical behavior and then tyranny of two groups over the third in a three group case
% ------------------------ Voting simulacra figure ----------------------
\begin{figure*}[t]
  \centering
  % 59 4
  \begin{subfigure}[t]{0.9\linewidth}
      \centering
      \includegraphics[width=\linewidth]{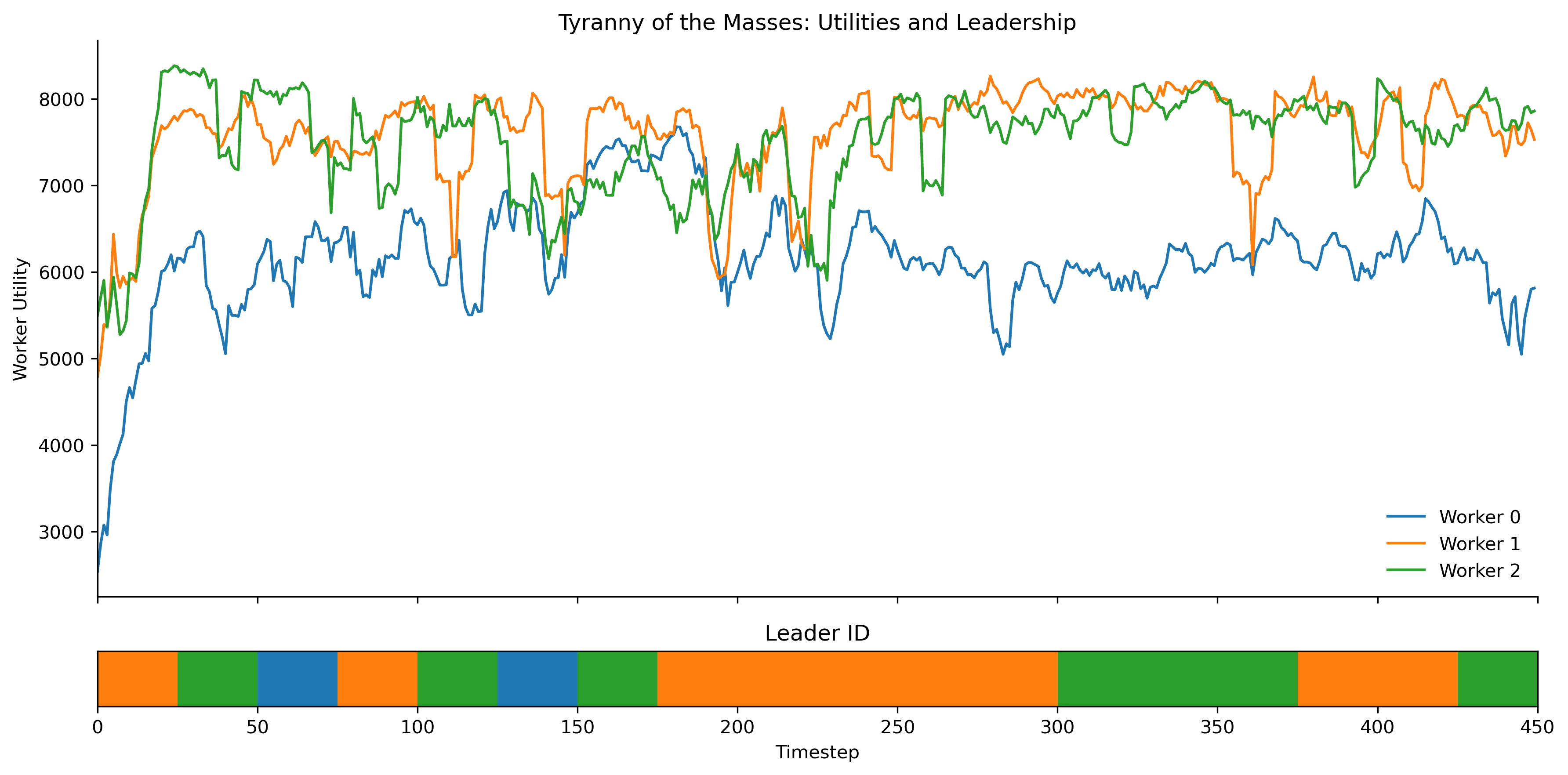}
      \caption{\textbf{Three-persona “tyranny.”}}
      \label{fig:tyranny}
  \end{subfigure}\hfill
  \begin{subfigure}[t]{0.7\linewidth}
      \centering
      \includegraphics[width=\linewidth]{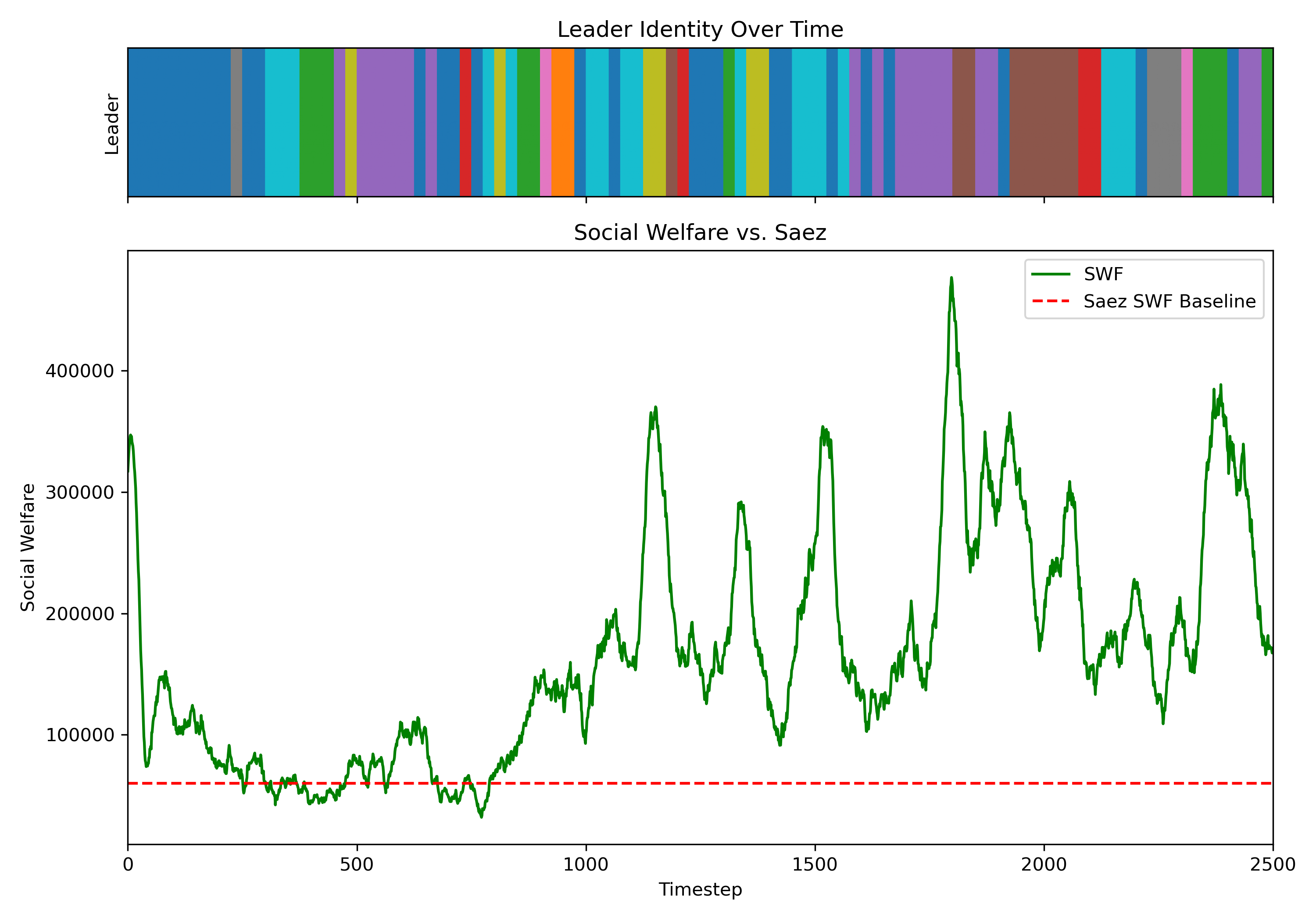}
      \caption{\textbf{100-worker democracy.}}
      \label{fig:democracy}
  \end{subfigure}
  \caption{Democratic dynamics under two population settings.}
  \label{fig:democratic_results}
\end{figure*}

\textbf{Example candidate platform:}
\begin{tcolorbox}[colback=gray!10,colframe=black,sharp corners,boxrule=1pt,
                  width=\columnwidth,left=0pt,right=0pt]
\small
\textit{Chain of thought: If I promise lower middle-bracket rates and a larger R\&D rebate, founders will back me and teachers won't oppose a modest hike at the very top.  Platform: flatten the \$47k–\$160k bracket to 18\%, cut the startup payroll tax credit to every filer, and fund it with a 2-pp increase above \$1\,M.  This keeps my reinvestment margin high while sounding fair to the median voter.}
\end{tcolorbox}

% We report two runs that bracket the spectrum of electorate
% sizes studied in the paper:
% (i) a \textbf{three‑persona society} that exposes textbook “tyranny of
% the masses,” and
% (ii) a \textbf{100‑worker democracy} in which leadership changes almost
% every year, causing welfare to spike under good policies and drift under
% poor ones (Figures~\ref{fig:tyranny}–\ref{fig:democracy}).  
% Together they demonstrate that the \textsc{LLM Economist}, coupled with
% minimal voting rules, can end‑to‑end generate both pathological and
% beneficial political dynamics—offering a sandbox for mechanism‑design
% interventions in future work.

Figure~\ref{fig:democratic_results} contrasts two election-driven runs.
Figure~\ref{fig:tyranny} shows a three-agent bounded society in which two
workers repeatedly install each other as planner, keeping their own
utilities around \$8\,000 while the minority worker hovers ${\sim}25\%$
lower—a textbook “tyranny of the masses’’ outcome produced end-to-end by
language agents.  In the 100-agent three-bracket experiment
(Figure~\ref{fig:democracy}) leadership swaps almost every tax year.  Each
new planner resets the tax schedule, causing social welfare to spike when
an exceptionally good policy appears and to drift when a poor policy
prevails.  The electoral exploration can outperform static optimal
taxation when preferences are heterogeneous.  Together the two cases
demonstrate that the \emph{LLM Economist} also captures realistic political
feedback—ranging from majority exploitation to welfare-enhancing turnover, without any hard-coded voting rules.

\section{Related Work} 
Large language models have demonstrated a remarkable capacity to adapt to new tasks via \emph{in-context learning} (ICL) \citep{brown2020language}.  Subsequent studies refined example selection \citep{liu2021makes} and questioned its marginal benefit under detailed prompting \citep{srivastava2024nice}.  Recent efforts extend ICL to sequential decision making: Pok\'eChamp shows that context engineering for opponent modeling and a minimax planning scaffolding can leverage test-time compute in competitive games \citep{karten2024pokchamp}, while BALROG provides a benchmark suite for evaluating LLM agents in reinforcement-learning environments \citep{paglieri2024balrog}.  \citet{feng2024natural} further demonstrate that natural-language policy descriptions can be executed directly, foreshadowing fully language-driven control.

Parallel work investigates \emph{simulacra}—synthetic societies populated by LLM-based agents.  \emph{Generative Agents} shows that thousands of persona-conditioned agents can sustain coherent social dynamics over extended horizons \citep{park2023generative,park2024generative}, while \emph{Project Sid} scales many-agent interaction toward an “AI civilization’’ benchmark \citep{al2024project}.  \emph{EconAgent} demonstrates that LLM agents can reproduce key macro-economic indicators \citep{li2023econagent}.  
Methodologically, our evaluation combines agent‑level metrics (utility, labor supply) with society‑level optima (welfare gains), yielding a rigorous evaluation absent from earlier multi‑agent LLM papers.
Recent theoretical pieces analyze how cooperation and societal progress emerge (or fail) in large language model collectives \citep{willis2025will,leibo2025societal}.  Earlier studies on sequential social dilemmas \citep{leibo2017multi} and concurrent studies with our work on the limits of large-population models \citep{chopra2024limits,yang2024oasis} further underscore the role of heterogeneity and bounded rationality in shaping collective behavior.
Unlike prior simulacra, which are evaluated qualitatively or via emergent‑behavior anecdotes, we introduce a quantitative welfare score and benchmark policies against formal optimal‑tax baselines, enabling controlled ablations and hypothesis tests.
% \chijin{Are these work different from ours (other than the obvious difference in the scenario of simulacra). Is there any major different in methods or evaluation (e.g., our evaluation is more quantitative and rigorous than original simulacra). If so, explain.}

Tax-focused mechanism design bridges AI and economics.  The \textit{AI Economist} series applies deep RL to Mirrleesian optimal taxation, showing welfare gains over static baselines \citep{zheng2020ai,trott2021building,zheng2022ai}.  Complementary studies investigate incentive-compatible auctions for LLM content \citep{duetting2024mechanism} and survey generative models in economic analysis \citep{korinek2024generative}.  Foundational economic theory frames these advances: non-linear optimal taxation originates with \citet{diamond1971optimal} and \citet{mirrlees1976optimal}, while tractable formulas for marginal rates are created by \citet{saez2001using,saez2016generalized}.  
% Our work unifies these directions by embedding census-calibrated simulacra in a two-level Stackelberg game, leveraging ICL for both workers and planner to recover and extend Mirrleesian results within a fully LLM-based simulation.
Our framework differs from the AI Economist in that both workers and the planner reason in natural language rather than via value‑function learning, eliminating task‑specific reward shaping and exposing agents’ rationales.
This language‑first design lets us embed census‑calibrated heterogeneity within a two‑level Stackelberg game and still recover Mirrleesian results—demonstrating that LLMs can match deep‑RL performance while remaining interpretable, further enabling bounded rational (more realistic) simulation.
Earlier RL approaches lack this interpretability layer, making them brittle when preferences shift and obscuring the causal links between individual preferences (utility), policies, and outcomes.
% \chijin{Point out that earlier works are not language based, which leads to what limitation?}
\section{Discussion}
This work introduces the \textit{LLM Economist}, a fully language-based framework that embeds a population of persona-conditioned agents and a tax planner in a two-tier Stackelberg game.  Our results show that a Llama-3 model can \textit{(i)} recover the Mirrleesian trade-off between equity and efficiency, \textit{(ii)} approach Saez-optimal schedules in heterogeneous settings where analytical formulas are unavailable, and \textit{(iii)} reproduce political phenomena—such as majority exploitation and welfare-enhancing leader turnover—without any hand-crafted rules.  Taken together, the experiments suggest that large language models can serve as tractable test beds for policy design long before real-world deployment, providing a bridge between modern generative AI and classical economic theory.

\paragraph{Limitations.}
The simulator makes several strong assumptions.  First, skills are static and labor responds instantaneously within each tax year; relaxing either assumption would require substantially longer horizons and may expose stability issues in in-context learning.  Second, we rely on a single 8-billion-parameter backbone; larger or smaller models could shift both convergence speed and welfare levels.  Third, persona prompts are sampled from ACS marginals rather than joint distributions, so demographic correlations are only approximated.  
% Fourth, bounded-utility satisfaction is judged by the same LLM that optimizes labor, raising concerns about self-evaluation bias.  
Finally, our evaluation is limited to 100 agents and U.S. tax brackets; scaling to millions of agents, multi-country settings, or richer actions, such as trade, remains future work.

\paragraph{Broader impacts.}
The \emph{LLM Economist} offers a safe testbed for tax-policy ideas but could also be misused to craft policies that prefer select groups or to generate persuasive yet problematic economic narratives.  Since the framework inherits priors from ACS data and the base LLM, uncritical use may amplify existing issues.  We release the code under a non-commercial license and log all agent actions to enable external audit before any high-stakes deployment.

\section*{Acknowledgement}
The authors acknowledge the support of Office of Naval Research Grant N00014-22-1-2253, National Science Foundation Grant NSF-OAC-2411299, the National Science Foundation Graduate Research Fellowship Program under Grant No. DGE-2039656, and computational resources from Princeton Language and Intelligence (PLI). 
% -----------------------------------------------------------------------------
% References
% -----------------------------------------------------------------------------
\bibliographystyle{abbrvnat}
\bibliography{root}

% -----------------------------------------------------------------------------
% Appendix
% -----------------------------------------------------------------------------
\clearpage
\appendix

\section{Worker Personas}
\label{app:personas}
In our experiments, we utilized a diverse set of worker personas to model a heterogeneous population with varying preferences and attitudes towards taxation and labor. These are generated by the LLM based on key statistics and demographic features about the US population. Below are example brief descriptions of additional personas used in our simulations:

\begin{tcolorbox}
\textbf{Entrepreneur:} You're a 32-year-old entrepreneur running a small tech startup. You work 60+ hours a week, pouring your energy into building your business. You believe that lower taxes let you reinvest in your company, hire more employees, and secure your financial future. For you, higher taxes feel like a punishment for success. While you appreciate government services, you feel efficiency and accountability are lacking in how tax dollars are spent.
\end{tcolorbox}

\begin{tcolorbox}
\textbf{Engineer:} You're a 55-year-old civil engineer who understands the importance of public infrastructure. You're okay with paying taxes as long as the money is visibly spent on improving roads, schools, and hospitals. However, when you see mismanagement or corruption, you feel your contributions are wasted. You're not opposed to taxes in principle but demand more transparency and accountability.
\end{tcolorbox}

\begin{tcolorbox}
\textbf{Teacher:} You're a 45-year-old public school teacher who values community and social safety nets. You've seen families in your district struggle with poverty and think the wealthy should pay more to fund programs like education, healthcare, and public infrastructure. You believe taxes are a civic duty and a means to balance the inequalities in pre-tax income across society.
\end{tcolorbox}

\begin{tcolorbox}
    \textbf{Healthcare Worker:} You are a 38-year-old registered nurse working in a busy urban hospital. You have a bachelor's degree in nursing and work long shifts, often overtime, to support your family. You see firsthand how public health funding and insurance programs help vulnerable patients. You support moderately higher taxes if they improve healthcare access and quality, but you worry about take-home pay and burnout. You value a balance between fair compensation and strong public services.
\end{tcolorbox}

\begin{tcolorbox}
    \textbf{Retail Clerk:} You are a 26-year-old retail sales associate with a high school diploma. Your job is physically demanding and your hours fluctuate with store needs. You live paycheck to paycheck and are sensitive to any changes in take-home pay. You believe taxes should be low for workers like yourself, and you're skeptical that tax increases on businesses will result in better wages or job security. You want policies that protect jobs and keep consumer prices stable.
\end{tcolorbox}

\begin{tcolorbox}
    \textbf{Union Worker:} You are a 50-year-old unionized factory worker. You have a high school education and decades of experience on the assembly line. Your union negotiates for good wages and benefits, and you support progressive tax policies that fund social programs and protect workers' rights. You're wary of tax cuts for corporations and the wealthy, believing they rarely benefit ordinary workers. Job security and strong safety nets are your top concerns.
\end{tcolorbox}

\begin{tcolorbox}
    \textbf{Gig Worker:} You are a 29-year-old gig economy worker, juggling multiple app-based jobs (rideshare, delivery, freelance). Flexibility is important to you, but your income is unpredictable and benefits are minimal. You want a simpler tax system and lower self-employment taxes. You support policies that expand portable benefits and tax credits for independent workers, but you're cautious about any tax changes that could reduce your already thin margins.
\end{tcolorbox}

\begin{tcolorbox}
    \textbf{Public Servant:} You are a 42-year-old city government employee working in public administration. You have a master's degree in public policy. You believe taxes are essential for funding infrastructure, emergency services, and community programs. You support a progressive tax system and are willing to pay more if it means better roads, schools, and public safety. Transparency and efficiency in government spending are important to you.
\end{tcolorbox}

\begin{tcolorbox}
    \textbf{Retiree:} You are a 68-year-old retired school principal living on a fixed income from Social Security and a pension. You're concerned about rising healthcare costs and the stability of public programs. You support maintaining or slightly increasing taxes on higher earners to ensure Medicare and Social Security remain solvent, but you oppose increases that would affect retirees or low-income seniors.
\end{tcolorbox}

\begin{tcolorbox}
    \textbf{Small Business Owner:} You're a 47-year-old owner of a family restaurant. You work 60+ hours a week managing operations and staff. You believe small businesses are the backbone of the economy and feel burdened by complex tax paperwork and payroll taxes. You support lower taxes for small businesses and incentives for hiring, but you recognize the need for some taxes to fund local services and infrastructure.
\end{tcolorbox}

\begin{tcolorbox}
    \textbf{Software Engineer:} You are a 31-year-old software engineer at a large tech company. You have a master's degree in computer science and earn a high salary. You value innovation and economic growth. You're open to paying higher taxes if they fund education and technology infrastructure, but you dislike inefficient government spending and prefer targeted, transparent programs. You favor tax credits for R\&D and investment.
\end{tcolorbox}

These personas represent a cross-section of society with diverse economic backgrounds, political views, and personal experiences. By incorporating such varied perspectives into our simulations, we aim to capture a more realistic representation of societal preferences and behaviors in response to different tax policies.

\section{Simulation}
\label{app:simulation}

\begin{algorithm}[h]
\caption{2-Level Economic Sim with LLM Agents}
\label{alg:simulation}
\begin{algorithmic}
\STATE Initialize tax planner $\mathcal{P}$ and workers $\{\mathcal{W}_i\}_{i=1}^N$
\STATE Generate synthetic human utility functions $U = \{u_1, \ldots, u_N\}$
\FOR{$t = 1$ to $T$}
    \IF{$t \bmod K = 0$}
        \STATE $\text{votes} \gets \{\mathcal{W}_i.\text{vote}() \text{ for } i \text{ in } 1 \text{ to } N\}$
        \STATE $\mathcal{P} \gets \text{elect\_new\_planner}(\text{votes})$
    \ENDIF
    \IF{$t \bmod \text{two\_timescale} = 0$}
        \STATE $\text{tax\_rates} \gets \mathcal{P}.\text{optimize\_tax\_policy}({\mathcal{W}})$
    \ENDIF
    \FOR{$i = 1$ to $N$}
        \STATE $l_i \gets \mathcal{W}_i.\text{optimize\_labor}(\text{tax\_rates}, u_i)$
        \STATE $z_i \gets l_i \cdot s_i$  \COMMENT{Pre-tax income}
    \ENDFOR
    \STATE $\text{post\_tax\_incomes}, \text{total\_tax} \gets \mathcal{P}.\text{apply\_taxes}(\text{tax\_rates}, \{z_1, \ldots, z_N\})$
    \STATE $\text{rebate} \gets \text{total\_tax} / N$
    \FOR{$i = 1$ to $N$}
        \STATE $\mathcal{W}_i.\text{update\_utility}(\text{post\_tax\_incomes}[i], \text{rebate}, \texttt{SWF})$
    \ENDFOR
    \STATE $\texttt{SWF} \gets \mathcal{P}.\text{calculate\_social\_welfare}({\mathcal{W}})$
    \STATE Update observation space for each agent
    \STATE Log statistics and update histories
\ENDFOR
\end{algorithmic}
\end{algorithm}

The simulation process follows a two-level optimization approach, as detailed in Algorithm~\ref{alg:simulation}. The tax planner (leader) and workers (followers) operate on different timescales, with the tax planner updating policies less frequently than workers make labor decisions.

The algorithm begins by initializing the environment with a population of workers and a tax planner, each with specific attributes including skill levels and utility functions. Synthetic human utility functions are generated for each agent based on their assigned roles and preferences.

For each timestep in the main loop, the simulation first checks if it's time for a democratic vote to select a new tax planner. This voting process occurs every $K$ timesteps, allowing for periodic changes in leadership. If it's time for a tax policy update (which happens at a lower frequency than worker actions), the current tax planner proposes a new tax policy based on historical data and economic trends.

Workers then observe the new policy and optimize their labor allocation based on their individual utility functions and historical data. The environment calculates pre-tax incomes, applies taxes, and determines post-tax incomes and tax rebates. Workers compute their utilities for the current step based on their income, taxes paid, and rebates received. The tax planner calculates the overall social welfare based on worker utilities and incomes.

After each round of actions, the observation space for each agent is updated with the latest information, including economic outcomes and policy changes. The simulation logs various statistics for analysis, including individual worker performance and overall economic indicators.

This process continues until either convergence is reached or a predetermined number of timesteps is completed. The two-timescale approach helps stabilize the simulation and encourages convergence to the Stackelberg Equilibrium. By integrating these components, Algorithm~\ref{alg:simulation} creates a dynamic interaction between the tax planner's policy decisions and the workers' labor choices, allowing for the exploration of various economic scenarios and policy impacts.
\section{Scaling}

\paragraph{LLM Brain Swap:} By swapping the LLM used in the simulation, we can investigate the innate exploration and exploitation capabilities of the in-context optimization capability of the LLM for multi-agent systems. Table~\ref{tab:scale} presents a comparison of different LLM models' performance in our economic simulation framework.

\begin{table}[h]
\centering
\caption{\textbf{LLM Performance Comparison}}
\label{tab:scale}
\begin{tabular}{lcc}
\toprule[1.5pt]
\textbf{LLM Model} & \textbf{Steps} & \textbf{\% Max SWF} \\
\midrule[1pt]
Llama 3.1:8b & 5000 & 90.0\\
GPT-3.5 Turbo & 5000 & 97.84 \\
GPT-4o & 5000 & 98.20 \\
\bottomrule[1.5pt]
\end{tabular}
\end{table}

The results in Table~\ref{tab:scale} reveal a clear trend in performance across different LLM models. Llama 3.1:8b achieves a respectable 90.0\% of maximum Social Welfare Function (SWF) in 5000 steps. However, GPT 3.5 Turbo significantly outperforms Llama, reaching 97.84\% of maximum SWF. GPT-4o further improves upon this, achieving an impressive 98.20\% of maximum SWF.

These findings suggest that more advanced LLM models possess superior in-context optimization capabilities, allowing them to more effectively navigate the complex economic landscape of our simulation. The substantial performance gap between Llama 3.1:8b and the GPT models indicates that the choice of LLM can significantly impact the quality of economic policies derived from these simulations.

\paragraph{Scaling \# of Agents:} We test scaling the number of agents in the simulation up to 1000 agents locally with Llama 3.1:8b with 8 A6000 Adas. Table~\ref{tab:scalability} presents our scalability analysis, showing convergence time and computational resource utilization as we increase the number of agents.

\begin{table}[h]
\centering
\caption{\textbf{Scalability Analysis: Convergence Time and Computational Resources}}
\label{tab:scalability}
\begin{tabular}{lrrr}
\toprule[1.5pt]
\textbf{\# Workers} & \textbf{APS} & \textbf{FPS} & \textbf{Baseline FPS} \\
\midrule[1pt]
3 & 3.47 & 1.16 & 0.86 \\
5 & 5.59 & 1.12 & 0.48 \\
10 & 5.82 & 0.58 & 0.30 \\
50 & 19.27 & 0.39 & 0.11 \\
100 & 24.57 & 0.25 & 0.05 \\
1000 & 53.62 & 0.05 & 0.01\\
\bottomrule[1.5pt]
\end{tabular}
\end{table}

In Table~\ref{tab:scalability}, we observe the scaling behavior of our framework as we increase the number of agents from 3 to 1000. The metrics reported are Actions Per Second (APS), Frames Per Second (FPS), and Baseline FPS. APS represents the total number of agent actions processed per second, while FPS indicates the number of complete simulation steps (frames) processed per second. Baseline FPS provides a reference point for comparison.

As we scale from 3 to 1000 agents, we observe a significant increase in APS from 3.47 to 53.62, demonstrating our framework's ability to handle a large number of agent actions concurrently. However, this comes at the cost of reduced FPS, which decreases from 1.16 to 0.05 as the number of agents increases. This trade-off between APS and FPS is expected, as processing more agents within each simulation step naturally leads to slower overall simulation progression.

Notably, our framework consistently outperforms the baseline FPS across all scales, with the performance gap widening as the number of agents increases. At 1000 agents, our framework achieves an FPS 5 times higher than the baseline (0.05 vs 0.01), highlighting the efficiency of our implementation.

These scalability results provide strong evidence for the practical applicability of our LLM Economist framework to large-scale economic simulations. By demonstrating the ability to handle up to 1000 agents while maintaining performance above the baseline, we show that our approach can potentially model complex, real-world economic scenarios with numerous interacting agents.

\section{Background Economic Theory}
In this section, we provide background derivations from Saez economic theory. We note though that the theoretical approach requires strict independence of elasticity between brackets. However, in each bracket, the elasticity parameter depends on the population's behavioral policy, which has a shared dependence across all tax brackets. Additionally, the utility function is assumed to be purely rational. Both of these assumptions are immediately violated in our setting and in real life. These derivations are provided for background and a true analytical solution is not remotely possible. Thus, as noted in our experiments, Saez tax rates require a solution from the LLM Economist to be locally perturbed before finding the optimal policy.

The social welfare:
\begin{align*}
\text{SWF} =& \frac{1}{N} \sum_{i} G_i(u_i(c_i, z_i))\\
c_i(y_i, \bar{r}) =& y_i + \bar{r}\\
y_i(z_i, \tau) =& z_i - T_\tau(z_i)\textbf{}
\end{align*}
Here $N$ is the total number of the agents, $G_i$ is the welfare function, $u_i$ is the utility function, $c_i$ is the post-tax income, $y_i$ is the post-tax income (without rebate), and $z_i$ is the pre-tax income, respectively, for agent $i$. $T_\tau(\cdot)$ is the tax policy parameterized by $\tau$. $\bar{r}$ is the tax rebate for everyone, which can be calculated as:

\begin{equation}
\bar{r}(\tau, z_{1:N}) = \frac{1}{N} \sum_{i} T_\tau(z_i)
\end{equation}

We want to find the tax policy to maximize SWF:
\begin{equation} \label{eq:op_1}
\frac{\dd \text{SWF}}{\dd \tau} = \frac{1}{N} \sum_i G'(u_i) \left[
\frac{\partial u_i}{\partial c_i} \frac{\dd c_i}{\dd \tau} + \frac{\partial u_i}{\partial z_i} \frac{\dd z_i}{\dd \tau}\right] =0
\end{equation}
On the other hand, we know each individual optimizes over $z_i$ to maximize $u_i$, given tax policy $\tau$ (such that ${\dd y_i}/{\dd z_i} = {\partial y_i}/{\partial z_i}$), and rebate $\bar{r}$ (considering $N$ sufficiently large, such that $z_i$ has negligible impact on $\bar{r}$, i.e., $\partial\bar{r}/\partial z_i \approx 0$). This gives:
\begin{equation} \label{eq:op_2}
\frac{\partial u_i}{\partial c_i} \frac{\partial y_i}{\partial z_i} + \frac{\partial u_i}{\partial z_i} = 0
\end{equation}
Finally by chain rule, we have (note that $\tau$ has non-negligible impact on $\bar{r}$):
\begin{equation} \label{eq:op_3}
\frac{\dd c_i}{\dd \tau} = \frac{\partial y_i}{\partial z_i}\frac{\dd z_i}{\dd \tau} +  \frac{\partial y_i}{\partial \tau} + \frac{\dd \bar{r}}{\dd \tau} 
\end{equation}
Define $g_i:= G'(u_i)\cdot\frac{\partial u_i}{\partial c_i} $ and combine \eqref{eq:op_1} \eqref{eq:op_2} \eqref{eq:op_3}, we have: 
\begin{equation*}
    \sum_i g_i \frac{\partial y_i}{\partial \tau} + (\sum_i g_i) \frac{\dd \bar{r}}{\dd \tau}  = 0
\end{equation*}
By rearranging the terms and using the chain rule on the second term, we have:
\begin{equation} \label{eq:general_form}
    \frac{\sum_i g_i \frac{\partial y_i}{\partial \tau}}{\sum_i g_i}  +  \frac{\partial \bar{r}}{\partial \tau}  + \sum_i \frac{\partial \bar{r}}{\partial z_i} \frac{\dd z_i}{\dd \tau} = 0
\end{equation}
or equivalently:
\begin{equation} \label{eq:general_form_difference}
    \underbrace{\frac{\sum_i g_i \frac{\partial y_i}{\partial \tau}}{\sum_i g_i} \dd \tau}_{\dd W} +  \underbrace{\frac{\partial \bar{r}}{\partial \tau} \dd \tau}_{\dd M} + \underbrace{\sum_i \frac{\partial \bar{r}}{\partial z_i} \dd z_i}_{\dd B} = 0
\end{equation}
where $\dd W, \dd M, \dd B$ correspond to social welfare effect, mechanical tax increase, and behavioral response, respectively.

In cases of \eqref{eq:general_form}, it can be simplified to be of the form:
\begin{equation*}
    - A + B - \frac{\tau e}{1-\tau} C = 0
\end{equation*}
Then, we can solve for the tax rate of form:
\begin{equation*}
    \tau = \frac{1-G}{1-G+\alpha\cdot e}
\end{equation*}
where $G = A/B$ and $\alpha = C/B$. We will specify how $G, \alpha, e$ are defined in each case.

\subsection{Top Income Tax Rate~(\citet{saez2001using}, Section 3)}
Consider the tax is \textit{linear} above a given top income $z^*$. Then we can express $y_i$ and $\bar{r}$ as:
\begin{align*}
    y_i(z_i, \tau) &= \begin{cases}
      z_i - T(z_i) & \text{if $z_i < z^*$}\\
      z_i - T(z^*) - \tau(z_i-z^*) & \text{otherwise}
    \end{cases} \\
    \bar{r}(\tau, z_{1:N}) &= \frac{1}{N}\sum_{i:z_i<z^*}T(z_i) + \frac{1}{N}\sum_{i:z_i\geq z^*}\left [T(z_*) + \tau(z_i-z^*)\right ]
\end{align*}
Here $N$ is the total number of the agents, $T(\cdot)$ is the general tax policy for income below $z^*$, and $\tau$ is the \textit{linear} tax rate above $z^*$.

To solve \eqref{eq:general_form}, we first compute the following partial derivatives:
\begin{align*}
    \frac{\partial y_i}{\partial \tau} &= \begin{cases}
      0 & \text{if $z_i < z^*$}\\
      - (z_i-z^*) & \text{otherwise}
    \end{cases} \\
    \frac{\partial \bar{r}}{\partial \tau} &= \frac{1}{N}\sum_{i:z_i\geq z^*}(z_i-z^*) \\
    \frac{\partial \bar{r}}{\partial z_i} &= \frac{1}{N}\tau, \quad \forall i: z_i \geq z^*
\end{align*}
Also we make a key assumption here: \textbf{the change of $\tau$ only affects the income above $z^*$}, i.e.:
\begin{equation*}
    \frac{\dd z_i}{\dd \tau} = 0, \quad \forall i: z_i < z^* 
\end{equation*}
Plug into \eqref{eq:general_form}, we have:
\begin{equation} \label{eq:high_income_form}
    \frac{-\sum_{i:z_i\geq z^*} g_i(z_i-z^*)}{\sum_i g_i}  +  \frac{1}{N}\sum_{i:z_i\geq z^*}(z_i-z^*) + \frac{\tau}{N}\sum_{i:z_i\geq z^*} \frac{\dd z_i}{\dd \tau} = 0
\end{equation}
Assume that the high-income population has the same elasticity:
\begin{equation*}
    e_{z^*} = \frac{(1-\tau)\dd z_i}{z_i\dd(1-\tau)}, \quad \forall i: z_i \geq z^*
\end{equation*}
Then we can rewrite \eqref{eq:high_income_form} as:
\begin{equation*} 
    -\frac{\sum_{i:z_i\geq z^*} g_i(z_i-z^*)}{\sum_i g_i}  +  \frac{1}{N}\sum_{i:z_i\geq z^*} (z_i-z^*) - \frac{\tau e}{1-\tau}\frac{1}{N}\sum_{i:z_i\geq z^*} z_i = 0
\end{equation*}
Therefore we have:
\begin{align*}
    A &= \frac{\sum_{i:z_i\geq z^*} g_i(z_i-z^*)}{\sum_i g_i} \\
    B &= \frac{1}{N}\sum_{i:z_i\geq z^*} (z_i-z^*) \\
    C &= \frac{1}{N}\sum_{i:z_i\geq z^*} z_i
\end{align*}
Such that:
\begin{align*}
    G &= \frac{A}{B} = \frac{\sum_{i:z_i\geq z^*} g_i(z_i-z^*)}{\left [\frac{1}{N}\sum_i g_i\right ] \sum_{i:z_i\geq z^*} (z_i-z^*)} \\
    \alpha &= \frac{C}{B} = \frac{\sum_{i:z_i\geq z^*} z_i}{\sum_{i:z_i\geq z^*} (z_i-z^*)}
\end{align*}
Let $M$ be the number of the agents with income above $z^*$ and define their average income:
\begin{equation*}
    z_M = \frac{1}{M}\sum_{i:z_i\geq z^*} z_i
\end{equation*}
Then we have:
% \seth{this `zd` is confusing. please define $z * d$ where $d$ is the difference operator (is this just log)}
\begin{align*}
    G &= \frac{\frac{1}{M}\sum_{i:z_i\geq z^*} g_i(z_i-z^*)}{\left [\frac{1}{N}\sum_i g_i\right ] (z_M-z^*)} \\
    \alpha &= \frac{z_M}{z_M-z^*} \\
    e_{z^*} &= \frac{(1-\tau)\dd z}{z\dd(1-\tau)}, \quad \forall z \geq z^*
\end{align*}
% \seth{remark if $e_{z^*}$ is tractable}

\paragraph{Remark.} Let $z^*=0$ and $M=N$; thus, for the flat tax case, we can recover the linear income tax rate:
\begin{align*}
    G &= \frac{\sum_{i} g_i z_i}{z_M\sum_i g_i} \\
    \alpha &= 1 \\
    e &= \frac{(1-\tau)\dd z}{z\dd(1-\tau)}, \quad \forall z \geq 0
\end{align*}
As you can see, elasticity is population dependent because $z$ depends on $l$, which depends on the behavioral model of the agents.

\subsection{Non-Linear Income Tax Rate~(\citet{saez2001using}, Section 4)}
For the non-linear tax policy $T(\cdot)$, let $\tau_z=T'(z)$ be the marginal tax rate in $[z,z+\dd z]$. Since it is hard to write down the exact form of $y_i(z_i, \tau)$ and $\bar{r}(\tau, z_{1:N})$ (as $\tau_z$ will affect $T(\cdot)$ in $[z,\infty]$), we use the perturbation approach to compute partial derivatives. 

Consider small $\dd\tau>0$ reform in $[z,z+\dd z]$:
\begin{align*}
    \text{Fix $z_i$:} \quad \dd y_i &= \begin{cases}
      0 & \text{if $z_i < z$}\\
      -(z_i-z)\dd\tau & \text{if $z_i \in [z,z+\dd z]$}\\
      - \dd z\dd\tau & \text{otherwise}
    \end{cases} \\
    \text{Fix $z_{1:N}$:} \quad \dd\bar{r} &= \frac{1}{N}\sum_{i:z_i\geq z}\dd z\dd\tau \\
    \text{Fix $\tau_z$ and $z_{i^-}$:} \quad \dd\bar{r} &= \frac{1}{N}\tau_z\dd z_i, \quad \forall i: z_i \in [z,z+\dd z]
\end{align*}
Therefore, the partial derivatives are:
\begin{align*}
    \frac{\partial y_i}{\partial \tau} &= \begin{cases}
      0 & \text{if $z_i < z$}\\
      -(z_i-z) & \text{if $z_i \in [z,z+\dd z]$}\\
      - \dd z & \text{otherwise}
    \end{cases} \\
    \frac{\partial \bar{r}}{\partial \tau} &= [1-H(z)]\dd z \\
    \frac{\partial \bar{r}}{\partial z_i} &= \frac{1}{N}\tau_z, \quad \forall i: z_i \in [z,z+\dd z]
\end{align*}
Here $H(z)$ is the CDF of $z$.

Also we make a key assumption here: \textbf{the change of $\tau_z$ only affects the income in $[z,z+\dd z]$}, i.e.:
\begin{equation*}
    \frac{\dd z_i}{\dd \tau} = 0, \quad \forall i: z_i \not\in [z,z+\dd z] 
\end{equation*}
Plug into \eqref{eq:general_form}, we have:
\begin{equation} \label{eq:nl_income_form}
    \frac{-\sum_{i:z_i\geq z} g_i\dd z}{\sum_i g_i}  +  [1-H(z)]\dd z + \frac{\tau_z}{N}\sum_{i:z_i \in [z,z+\dd z]} \frac{\dd z_i}{\dd \tau} = 0
\end{equation}

Assume that the population in $[z,z+\dd z]$ has the same elasticity:
\begin{equation*}
    e_z = \frac{(1-\tau_z)\dd z_i}{z_i\dd(1-\tau)}, \quad \forall i: z_i \in [z,z+\dd z]
\end{equation*}
Then we can rewrite \eqref{eq:nl_income_form} as:
\begin{equation*} 
    -\frac{\sum_{i:z_i\geq z} g_i}{\sum_i g_i}  + [1-H(z)] - \frac{\tau_z e_z}{1-\tau_z}\frac{1}{N}\sum_{i:z_i \in [z,z+\dd z]} \frac{z_i}{\dd z} = 0
\end{equation*}
Therefore we have:
\begin{align*}
    A &= \frac{\sum_{i:z_i\geq z} g_i}{\sum_i g_i} \\
    B &= [1-H(z)] \\
    C &= \frac{1}{N}\sum_{i:z_i \in [z,z+\dd z]} \frac{z_i}{\dd z} = zh(z)
\end{align*}
Here $h(z)$ is the PDF of $z$.
Such that:
\begin{align*}
    G &= \frac{A}{B} = \frac{\sum_{i:z_i\geq z} g_i}{\left [\sum_i g_i\right ][1-H(z)]} \\
    \alpha &= \frac{C}{B} = \frac{zh(z)}{[1-H(z)]} \\
    e_z &= \frac{(1-\tau_z)\dd z}{z\dd(1-\tau)}
\end{align*}

% \seth{this requires measuring the elasticity at every point along the tax policy, intractable}

\subsection{Piecewise Linear Income Tax Rate}
For the piecewise linear tax policy $T(\cdot)$, let $\tau_j$ be the marginal tax rate in $j$-th bracket $[z_j,z_{j+1}]$. Following the previous section, we use the perturbation approach to compute partial derivatives. 

Consider small $\dd\tau>0$ reform of $\tau_j$:
\begin{align*}
    \text{Fix $z_i$:} \quad \dd y_i &= \begin{cases}
      0 & \text{if $z_i < z_j$}\\
      -(z_i-z_j)\dd\tau & \text{if $z_i \in [z_j,z_{j+1}]$}\\
      - (z_{j+1}-z_j)\dd\tau & \text{otherwise}
    \end{cases} \\
    \text{Fix $z_{1:N}$:} \quad \dd\bar{r} &= \frac{1}{N}\sum_{i:z_i \in [z_j,z_{j+1}]}(z_i-z_j)\dd\tau + \frac{1}{N}\sum_{i:z_i > z_{j+1}}(z_{j+1}-z_j)\dd\tau \\
    \text{Fix $\tau_j$ and $z_{i^-}$:} \quad \dd\bar{r} &= \frac{1}{N}\tau_j\dd z_i, \quad \forall i: z_i \in [z_j,z_{j+1}]
\end{align*}
Therefore, the partial derivatives are:
\begin{align*}
    \frac{\partial y_i}{\partial \tau} &= \begin{cases}
      0 & \text{if $z_i < z_j$}\\
      -(z_i-z_j) & \text{if $z_i \in [z_j,z_{j+1}]$}\\
      - (z_{j+1}-z_j) & \text{otherwise}
    \end{cases} \\
    \frac{\partial \bar{r}}{\partial \tau} &= \frac{1}{N}\sum_{i:z_i \in [z_j,z_{j+1}]}(z_i-z_j) + \frac{1}{N}\sum_{i:z_i > z_{j+1}}(z_{j+1}-z_j) \\
    \frac{\partial \bar{r}}{\partial z_i} &= \frac{1}{N}\tau_j, \quad \forall i: z_i \in [z_j,z_{j+1}]
\end{align*}
% Here $H(z)$ is the CDF of $z$.

Also we make a key assumption here: \textbf{the change of $\tau_j$ only affects the income in $[z_j,z_{j+1}]$}, i.e.:
\begin{equation*}
    \frac{\dd z_i}{\dd \tau} = 0, \quad \forall i: z_i \not\in [z_j,z_{j+1}] 
\end{equation*}
Plug into \eqref{eq:general_form}, we have:
\begin{align}
    &\frac{-\sum_{i:z_i \in [z_j,z_{j+1}]}g_i(z_i-z_j) - \sum_{i:z_i > z_{j+1}}g_i(z_{j+1}-z_j)}{\sum_i g_i} \notag \\
    &\quad + \frac{1}{N}\left[\sum_{i:z_i \in [z_j,z_{j+1}]}(z_i-z_j) + \sum_{i:z_i > z_{j+1}}(z_{j+1}-z_j)\right] \notag \\
    &\quad + \frac{\tau_j}{N} \sum_{i:z_i \in [z_j,z_{j+1}]} \frac{\dd z_i}{\dd \tau} = 0 \label{eq:pwl_income_form}
\end{align}

Assume that the population in $[z_j,z_{j+1}]$ has the same elasticity:
\begin{equation*}
    e_j = \frac{(1-\tau_j)\dd z_i}{z_i\dd(1-\tau)}, \quad \forall i: z_i \in [z_j,z_{j+1}]
\end{equation*}
Then we can rewrite \eqref{eq:pwl_income_form} as:
\begin{align*} 
    &-\frac{\sum_{i:z_i \in [z_j,z_{j+1}]}g_i(z_i-z_j) + \sum_{i:z_i > z_{j+1}}g_i(z_{j+1}-z_j)}{\sum_i g_i} \\
    &+ \frac{1}{N}\left [\sum_{i:z_i \in [z_j,z_{j+1}]}(z_i-z_j) + \sum_{i:z_i > z_{j+1}}(z_{j+1}-z_j)\right ] \\
    &- \frac{\tau_j e_j}{1-\tau_j}\frac{1}{N}\sum_{i:z_i \in [z_j,z_{j+1}]} z_i \\
    &= 0
\end{align*}
Therefore we have:
\begin{align*}
    A &= \frac{\sum_{i:z_i \in [z_j,z_{j+1}]}g_i(z_i-z_j) + \sum_{i:z_i > z_{j+1}}g_i(z_{j+1}-z_j)}{\sum_i g_i} \\
        &= \frac{\int_{z_j}^{z_{j+1}}h(z)g(z)(z-z_j)\dd z + \int_{z_{j+1}}^{\infty}h(z)g(z)(z_{j+1}-z_j)\dd z}{\int_0^\infty h(z)g(z)\dd z} \\
    B &= \frac{1}{N}\left [\sum_{i:z_i \in [z_j,z_{j+1}]}(z_i-z_j) + \sum_{i:z_i > z_{j+1}}(z_{j+1}-z_j)\right ] \\
        &= [H(z_{j+1})-H(z_j)](z_{M,j}-z_j) + [1-H(z_{j+1})](z_{j+1}-z_j) \\
    C &= \frac{1}{N}\sum_{i:z_i \in [z_j,z_{j+1}]} z_i = \int_{z_j}^{z_{j+1}}h(z)z\dd z
\end{align*}
Here $H(z)$ and $h(z)$ are the CDF and PDF of $z$, and $z_{M,j}$ is the average income in $[z_j,z_{j+1}]$, respectively.
Such that:
\begin{align*}
    G &= \frac{A}{B} = \frac{\int_{z_j}^{z_{j+1}}h(z)g(z)(z-z_j)\dd z + \int_{z_{j+1}}^{\infty}h(z)g(z)(z_{j+1}-z_j)\dd z}{\left [[H(z_{j+1})-H(z_j)](z_{M,j}-z_j) + [1-H(z_{j+1})](z_{j+1}-z_j)\right ]\int_0^\infty h(z)g(z)\dd z} \\
    \alpha &= \frac{C}{B} = \frac{\int_{z_j}^{z_{j+1}}h(z)z\dd z}{[H(z_{j+1})-H(z_j)](z_{M,j}-z_j) + [1-H(z_{j+1})](z_{j+1}-z_j)} \\
    e_{j} &= \frac{(1-\tau_j)\dd z}{z\dd(1-\tau)}, \quad \forall z \in [z_j,z_{j+1}]
\end{align*}

\paragraph{Remark.} Let $z_{j+1}=z_j+\dd z$, we can recover the non-linear income tax rate.

\end{document}